\documentclass{aa}  
\usepackage[utf8]{inputenc}
\usepackage{natbib}
\usepackage{graphicx}
\usepackage{wrapfig}
\usepackage{textcomp, gensymb}
\usepackage{amsmath}
\usepackage{fancyhdr}
\usepackage{booktabs,caption}
\usepackage[flushleft]{threeparttable}
\usepackage{longtable}
\usepackage{xcolor}
\usepackage[version=4]{mhchem} 
\usepackage{color}
\usepackage{subcaption}
\usepackage{graphicx}

\newcommand{\hii}{H{\sc ii}}
\newcommand{\hi}{H{\sc i}}
\newcommand{\kms}{km s$^{-1}$}
\usepackage{txfonts}
\makeatletter
\renewcommand*\aa@pageof{, page \thepage{} of \pageref*{LastPage}}
\makeatother
\usepackage{hyperref}

\begin{document}

\title{The Cygnus Allscale Survey of Chemistry and Dynamical Environments: CASCADE}
\subtitle{III. The large scale distribution of \ce{DCO+}, DNC and DCN in the DR21 filament}

   \author{I. Barlach Christensen\thanks{Member of the International Max Planck Research School (IMPRS) for Astronomy and Astrophysics at the Universities of Bonn and Cologne}
          \inst{1}
          \and
          F. Wyrowski \inst{1}
          \and
          V. S. Veena \inst{1,3}
          \and
          H. Beuther \inst{2}
          \and
          D. Semenov \inst{2}
          \and
          K. M. Menten \inst{1}
          \and
          A. M. Jacob \inst{1, 4}
          \and
          W.-J. Kim \inst{3}
          \and
          N. Cunningham \inst{8}
          \and
          C. Gieser \inst{5}
          \and
          A. Hacar \inst{7}
          \and
          S. Li \inst{2}
          \and
          N. Schneider \inst{3}
          \and
          I. Skretas \inst{1}
          \and 
          J. M. Winters \inst{6}
          }

   \institute{Max-Planck-Institut f\"ur Radioastronomie, 
            Auf dem H\"ugel 69, 53121 Bonn, Germany\\
             \email{ibarlach@mpifr-bonn.mpg.de}
             \and
             Max Planck Institut f\"ur Astronomie, 
             K\"onigstuhl 17, 69117 Heidelberg, Germany
             \and
             Physikalisches Institut der Universit\"at zu K\"oln, Z\"ulpicher Str. 77, 50937 K\"oln, Germany
             \and
             William H. Miller III Department of Physics \& Astronomy, Johns Hopkins University, 3400 North Charles Street, Baltimore, MD 21218, USA
             \and
            Max-Planck-Institut f\"ur Extraterrestrische Physik, Giessenbachstrasse 1, 85748 Garching, Germany
            \and
            Institut de Radioastronomie Milli\'etrique (IRAM), 300 rue de la Piscine, Domaine Universitaire, 38406 St. Martin d'H\`eres, France
            \and
            Department of Astrophysics, University of Vienna, T\"urkenschanzstrasse 17, 1180 Vienna, Austria
            \and
            SKA Observatory, Jodrell Bank, Lower Withington, Macclesfield SK11 9FT, United Kingdom
            }

   \date{Received March 25, 2024; accepted June 5, 2024}

\abstract
    {Deuterated molecules and their molecular D/H-ratios ($R_D$(D)) are important diagnostic tools to study the physical conditions of star-forming regions. The degree of deuteration, $R_D$(D), can be significantly enhanced over the elemental D/H-ratio depending on physical parameters such as temperature, density, and ionization fraction.}
    {Within the \textbf{C}ygnus \textbf{A}llscale \textbf{S}urvey of \textbf{C}hemistry \textbf{a}nd \textbf{D}ynamical \textbf{E}nvironments (CASCADE), we aim to explore the large-scale distribution of deuterated molecules in the nearby ($d \sim 1.5$ kpc) Cygnus-X region, a giant molecular cloud complex that hosts multiple sites of high mass star formation. We focus on the analysis of large-scale structures of deuterated molecules in the filamentary region hosting the prominent \hii~region DR21 and DR21(OH), a molecular hot core that is at an earlier evolutionary state.}
    {The DR21 filament has been imaged using the IRAM 30-m telescope in a variety of deuterated molecules and transitions.
    Here we discuss the \ce{HCO+}, HNC and HCN molecules and their deuterated isotopologues \ce{DCO+}, DNC and DCN, and their observed line emissions at 3.6, 2, and 1.3-mm.}
    {The spatial distributions of integrated line emissions from \ce{DCO+}, DNC, and DCN reveal morphological differences. Notably, \ce{DCO+} displays the most extended emission, characterized by several prominent peaks. Likewise, DNC exhibits multiple peaks, although its emission appears less extended compared to \ce{DCO+}. In contrast to the extended emission of \ce{DCO+} and DNC, DCN appears the least extended, with distinct peaks.
    Focusing only on the regions where all three molecules are observed, the mean deuteration ratios for each species, $R_D$, are 0.01 for both DNC and DCN, and $=0.005$ for \ce{DCO+}, respectively.
    Anti-correlations are found with deuterated molecules and dust temperature or $N$(\ce{H2}).}
    {The strongest anti-correlation is found with $R_D$(\ce{DCO+}) and $N$(\ce{H2}), with a Pearson correlation coefficient of $\rho = -0.74$.
    We analyze the SiO emission as a tracer for shocks and the $N$(HCO)/$N$(\ce{H^13CO+}) as a tracer for increased photodissociation by UV.
    The anti-correlation of $R_D$(\ce{DCO+}) and $N$(\ce{H2}) is suggested to be a result of a combination of an increased photodissociation degree and shocks.
    A strong positive correlation between the ratio of integrated intensities of DCN and DNC with their $^{13}$C-isotopologues, are found in high column density regions.
    The positive relationship between the ratios implies that the D-isotopologue of the isomers could potentially serve as a tracer for the kinetic gas temperature.}

\maketitle

\section{Introduction}
\label{Sec:Introduction}
While an overall understanding has emerged that star formation occurs in cool, dense molecular clouds, the physical and chemical conditions during the formation of high-mass stars are still poorly understood. The chemistry of these clouds 
evolves in complexity, heavily depending on the underlying and evolving physical conditions (see, for example, \citealt{McKee2002,McKee2003,Girichidis2020, Tielens2021}). Conversely, the presence of molecules, primarily CO, and also dust grains affects the physical conditions and the evolution of the clouds, as these coolants are important to allow their contraction \citep{Hocuk2014}.

In the interstellar medium (ISM), the elemental abundance ratio of deuterium to hydrogen (D/H) is observed to be approximately $10^{-5}$ \citep{Linsky1998, Oliveira2003, Cooke2018}. 
The earliest phases of star-formation begin when the cloud is cold ($T <$ 30 K) and dense ($n > 10^{5}$ cm$^{-3}$) \citep[e.g.][]{Fontani2011}.
During this phase, deuterium chemistry begins via ion-molecule reaction involving \ce{H3+} and HD:
\begin{equation}
  \ce{H3+} + \ce{HD} \rightarrow \ce{H2D+} + \ce{H2} + \Delta E\,(230 \text{ K})
  \label{eq:lowtempdeut}
\end{equation}
enabling the transfer of deuterium from its main reservoir \ce{HD} to other species \citep{Watson1974,Wootten1987}. 
Additionally, depending on the spin-state of \ce{H3+}, and consequently of the formed \ce{H2D+}, the reverse reaction in Eq. \ref{eq:lowtempdeut} is enhanced \citep{Flower2006}.
During the freeze-out onto dust grains, the number of heavy neutral molecules in the gas phase, such as \ce{CO}, decreases. 
This decreases the destruction of \ce{H2D+} \citep{Pillai12, Sabatini2020}. 
Therefore, low temperatures lead to an accumulation of \ce{H2D+}. Reaction \ref{eq:lowtempdeut} is the first formation route for heavier deuterated molecules, leading to a higher molecular XD/XH ratio ($\equiv R_D$(XD), with XD and XH the abundance of deuterated and H-isotopolog, respectively) that can reach up to a few percent \citep{Roberts00, Roberts04} in dense, molecular clouds. 
Simple gas phase chemistry predicts equilibrium values of $R_D$(XD) for simple molecules to be of the order of several times 10\,\% for $R_D$(\ce{DCO+}) \citep{Roberts2000}. 
At higher temperatures ($T \sim 30 - 80$ K), deuteration mainly occurs through the light hydrocarbons, i.e., \ce{CH2D+} and \ce{C2HD+} \citep{Millar1989, Albertsson2013}. 
Furthermore, deuterated molecules can  also be formed on dust grain mantles where D-atoms react with ice. 
As the temperature drops below $\sim 20$ K, the flow of atomic D onto dust grains increases, and consequently the rate of grain-surface reactions with H and D increases \citep{Albertsson2013}. This can result in high XD/XH ratios in complex organic molecules (COMs; C-bearing molecules containing more than 6 atoms) of up to a few times 10\% in extreme cases, leading even to multiply deuterated COMs \citep{Parise2002}.
By studying the deuterated molecules and their main isotopologs, we can gain insight into the physical conditions at early-stage star formation, such as the gas density, temperature, and ionization fraction \citep{Herbst1982,Caselli1998,Favre2015,Gerner2015}.

In this paper, three deuterated molecules are of particular interest: \ce{DCO+}, DNC, and DCN. 
\ce{DCO+} is efficiently formed through \ce{H2D+} reacting with CO in the low-temperature regime \citep{Albertsson2013}.
An additional formation pathway for \ce{DCO+} in the higher temperature regime ($T > 50$ K) is thought to occur from the light hydrocarbon, \ce{CH2D+} \citep{Favre2015}. 
The column density ratio $R_D$(\ce{DCO+}) also depends on the ionization fraction \citep{Caselli1998}, where a decreasing ratio can suggest a higher ionization fraction \citep{Favre2015}. 
DNC and DCN show dissimilar behavior in their formation pathways. 
DNC is formed from \ce{HCND+}, via \ce{H2D+}, reacting with CN or HCN in the lower temperature regime ($\sim 20-30$~K) or through \ce{CH2D+} reacting with N in the higher temperature regime \citep{Turner2001,Albertsson2013}. 
On the other hand, DCN is formed primarily ($\sim 66$\%) through \ce{CH2D+} via \ce{CHD} or \ce{CH2D} reacting with N in conditions typical of translucent and dark clouds \citep{Turner2001}. A significant, but less efficient formation pathway of DCN is through \ce{DCO+} reacting with HCN or HNC forming \ce{HDCN+} or \ce{DCNH+}, respectively. This reaction becomes more efficient with increasing temperature, forming both DNC and DCN \citep{Albertsson2013}. Furthermore, the formation of DNC from HCN is more efficient than the formation of DCN from HNC \citep{Turner2001}. The abundances of both precursors \ce{H2D+} and \ce{CH2D+} are temperature dependent \citep{Albertsson2013}, which leads to the increasingly efficient production of DCN with temperature. This means that the formation of DNC is more efficient at lower temperatures ($\sim 30$ K) than that of DCN. As the temperature increases, the efficiency with which DCN is formed increases, surpassing the production of DNC at higher temperatures ($\sim 80$ K). 

\cite{Gerner2015} investigated the behavior of deuterated molecules' abundances in high-mass star-forming regions, which they divided into four different evolutionary stages, namely infrared dark clouds (IRDC), high-mass protostellar objects (HMPO), hot molecular cores (HMC), and ultra-compact \hii-regions (UC\hii), evolving as IRDC $\rightarrow$ HMPO $\rightarrow$ HMC $\rightarrow$ UC\hii. They show that the column densities, $N$, of \ce{DCO+} and DNC decrease with progressing evolutionary stage of a clump, whereas $N$(DCN) peaks at the HMC-stage \citep{Gerner2015}. 
In addition, they find the $R_D$(\ce{DNC}) and $R_D$(DCN) deuteration ratios show a range of values, with an average of 0.02 and 0.005, respectively \citep{Gerner2015}. In the same sources, $R_D$(\ce{DCO+}) was measured to be an order of magnitude lower with an average ratio of 0.0025 \citep{Gerner2015}.
Toward two dark cloud sources, L134N and TMC-1, $R_D$(\ce{DCO+}) was found to be 0.18 and 0.02, respectively \citep{Tine2000}. 
More recent observations toward the distant \hii-region W51 star-forming region showed a core, on the same spatial scales, with a comparable ratio of 0.02 \citep{Vastel2017}.

The prominent, nearby giant molecular cloud/star formation complex Cygnus-X is a prime target for studying the early stages of massive star formation. It has a rich environment which contains recent and on-going star formation activities. These are, OB-associations, Wolf-Rayet stars, numerous \hii-regions,
and dense molecular star-forming clumps \citep{Schneider2006, Motte07, Reipurth2008}. The most active and dense region within the Cygnus-X region is a filament that contains, near its southern end, the compact \hii-region DR21 Main \citep{Downes1966,Immer2014,Schneider10, Hennemann}, which has a rich molecular envelope with a photodissociation interface \citep{Ossenkopf2010}.
The filament is proposed to be the result of interacting \hi~flows and molecular clouds \citep{Schneider2023, Bonne2023}.
The DR21 region is located at a distance of 1.5\,kpc \citep{Rygl}, and its dense filamentary structure, also called the "DR21 ridge", harbors several dense cores with a gradient of evolutionary stages \citep{Hennemann}. 
Further north, a famous active \ce{OH}, \ce{H2O}, and  \ce{CH3OH} maser source, the dense protocluster/hot molecular core DR21(OH) \citep{Mangum1992} is found and a colder region of in-falling gas is located in the northernmost part of the ridge \citep{Schneider10, Hennemann, Koley2021}. Previous studies of \ce{H2D+} north of DR21(OH) using the James Clerk Maxwell Telescope (JCMT) have shown the emission to probe a large-scale cold, quiescent gas component \citep{Pillai12}, hinting also at the presence of cold gas pockets nearby, only slightly offset from the DR21 filament.

The \textbf{C}ygnus-X \textbf{A}llscale \textbf{S}urvey of \textbf{C}hemistry \textbf{A}nd \textbf{D}ynamical \textbf{E}nvironments (CASCADE; \citealt{Beuther2022}) is a large scale survey of the Cygnus-X region carried out with the NOrthern Extended Millimeter Array (NOEMA) and the Institut de Radioastronomie Millim\'etrique (IRAM) 30-m telescope as a Max Planck Institute and IRAM Observatory large-program (MIOP). The objective is to connect star formation properties at large scales probed with the IRAM 30-m telescope to small scale dense cores observed by the NOEMA interferometer. In its lower frequency range of 70--80 GHz, the project covers the ground state level transitions of several key deuterated species, \ce{DCO+}, DNC, and DCN (see Table~\ref{tab:Deuterated}). Higher rotational transitions of the deuterated species in the 1.3-mm and 2-mm wavelength bands were performed with follow-up observations once again using the IRAM 30-m telescope to constrain the line excitation temperatures that are used for determining reliable molecular column densities.

The paper is organized as follows: in Sect.~\ref{Sec:Observations}, we present the observations and the data reduction; in Sect.~\ref{Sec:Results} we show the integrated intensities and averaged spectra of deuterated molecules in the DR21 filament, and describe the calculation of molecular column densities. In Sect.~\ref{Sec:Analysis} and \ref{Sec:Discussion}, we analyze and discuss the molecular D/H ratios and their behavior with dust temperature and \ce{H2} column density.
Finally, in Sect.~\ref{sec:Conclusions}, we present our conclusions. 

\section{Observations}
\begin{figure*}[htpb!]
  \centering
  \sidecaption
  \includegraphics[width=0.66\linewidth]{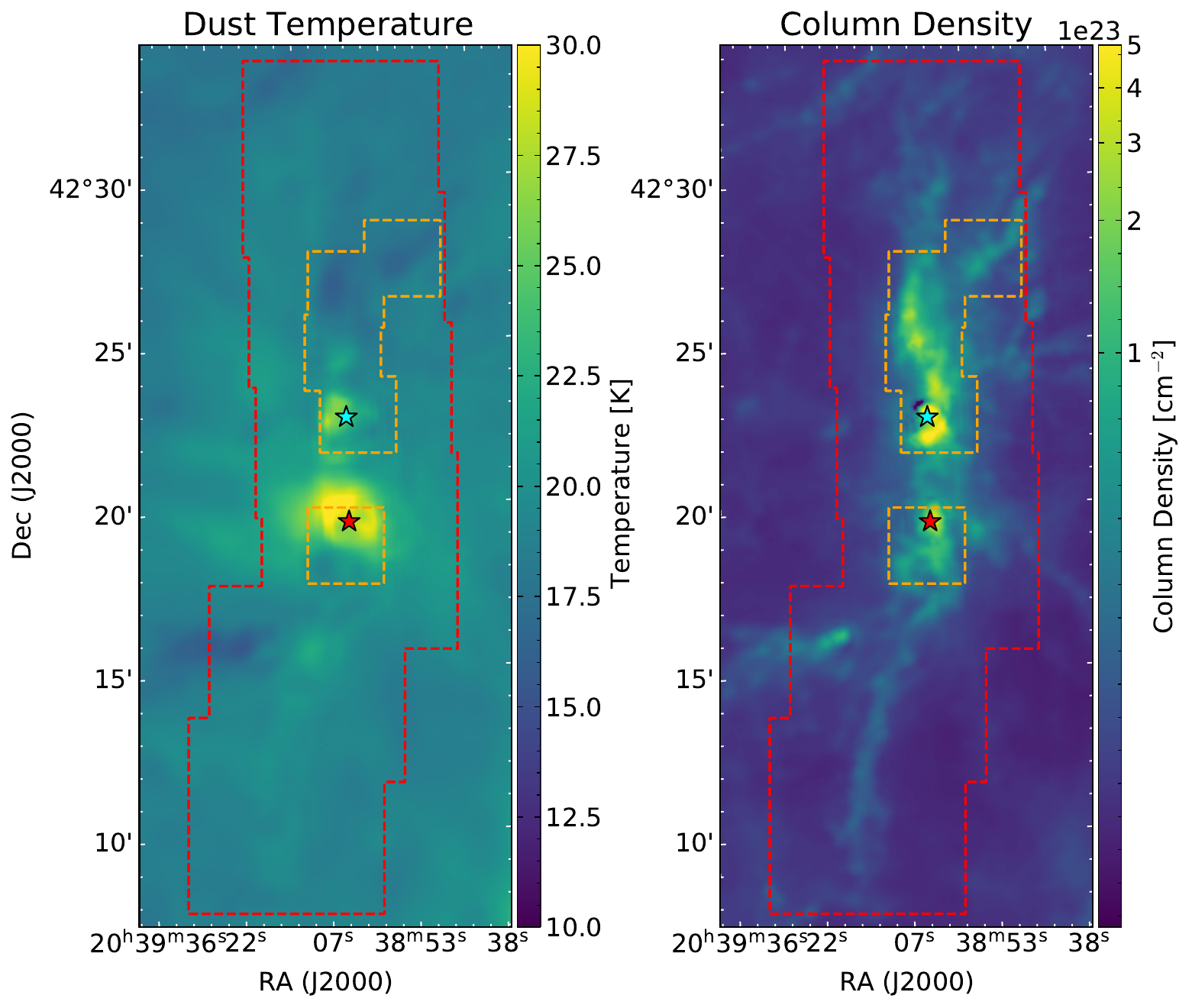}
  \caption{Maps of dust temperature (left) and \ce{H2} column density (right) of the DR21 filament taken determined by the HOBYS program \citep{Motte2010}. 
  The stars in both maps indicate the positions of DR21 Main (red) and DR21(OH) (blue). The red and orange dashed lines show the coverage of CASCADE and follow-up observations, respectively.} 
  \label{fig:DR21_maps}
\end{figure*}
\label{Sec:Observations}
We carried out On-The-Fly (OTF) mapping observations (Project ID: 145-19 and 031-20; PIs: F. Wyrowski and H. Beuther) at 4 millimeter (mm) wavelength using the IRAM 30-m telescope.
To obtain a large bandwidth, covering all of the deuterated molecules of interest, the mappings were performed  using the Eight MIxer Receiver (EMIR)\footnote{\url{http://www.iram.es/IRAMES/mainWiki/EmirforAstronomers}}  \citep{EMIR} and using the Fourier Transform Spectrometer backend with 195 \,kHz channel spacing (FTS200). A bandwidth of 8 GHz per sideband (in total, a frequency coverage of 16\,GHz) with EMIR E90 in both polarization was covered. At 72 GHz, the spectral channel resolution of 195 kHz is equivalent to a velocity resolution of $0.73$ \kms (smoothed to a spectral resolution of $\mathrm{d}\upsilon = 0.8$ \kms) and the  half power beam width (HPBW) of the IRAM 30-m telescope is $34\arcsec$.\footnote{This telescope's HPBW (in arcseconds) is well represented by 2460/$\nu$(GHz); see \url{https://publicwiki.iram.es/Iram30mEfficiencies}.}

Observations of the DR21 filament were carried out on 2020 May 21 and 22. 
The regions with the highest \ce{H2} column densities \citep{Motte2010} were covered in 6 tiles.
Figure~\ref{fig:DR21_maps} shows the DR21 coverage of CASCADE in red dashed lines.  
Each tile, of size  $\sim 306\arcsec \times 306 \arcsec$, is observed with the OTF mapping technique, in two scanning directions, in right ascension (R.A.) and in declination (Dec.) with a scanning frequency of $\sim 17\rlap{.}\arcsec5$ s$^{-1}$ and a dump time of $\sim 0.5$ s.
This results in an oversampling of the beam by a factor $>3$. 
For each mapping, two temperature load calibration observations, each of 60s duration, were done. In order to avoid edge-effects, tiles slightly overlap along the filament. 

\begin{table*}
  \centering
    \caption{Frequencies and spectroscopic properties of the deuterated molecules and their corresponding $^{13}$C-isotopologs presented in this study alongside the spectral line properties obtained by averaging over the entire DR21 filament are presented.} 
  \begin{tabular}{lrrrrcrcrrr}  \hline \hline \\ [-0.1mm]
  Species &  Transition & \multicolumn{1}{c}{Frequency}  & \multicolumn{1}{c}{$E_{\rm up}/k$} & \multicolumn{1}{c}{$A_{\rm ij}$} & \multicolumn{1}{c}{$g_{\rm up}$} & \multicolumn{1}{c}{$n_{\rm crit}$} & \multicolumn{1}{c}{Database} & \multicolumn{1}{c}{$T_{\text{MB}}$} & \multicolumn{1}{c}{$\upsilon_{\rm Peak}$} & \multicolumn{1}{c}{FWHM} \\ 
   &  & \multicolumn{1}{c}{[MHz]} & \multicolumn{1}{c}{[K]} & \multicolumn{1}{c}{[s$^{-1}$]} & & \multicolumn{1}{c}{[cm$^{-3}$]}  & & \multicolumn{1}{c}{[K]} & \multicolumn{1}{c}{[\kms]} & \multicolumn{1}{c}{[\kms]}
  \\ \hline 
  DCO$^+$ & $J=1 - 0$ & 72039.312 &  3.46 & 2.2$(-5)$ & 3 & 4.5(4) & CDMS & 0.08 $\pm$ 0.01 & $-$2.7 $\pm 0.8$ & 3.0 $\pm$ 0.1 \\ 
  DCO$^+$ & $J= 3 - 2$ & 216112.582 &  20.74 & 7.1$(-4)$ & 7 & 1.0(6) & CDMS & 0.14 $\pm$ 0.01 & $-$3.2 $\pm 0.8$ & 3.4 $\pm$ 0.1 \\  
  H$^{13}$CO$^+$ & $J= 1 - 0$ & 86754.288 &  4.16 & 3.9$(-5)$ & 3 &  4.5(4) & CDMS & 0.21 $\pm$ 0.01 & $-$3.0 $\pm 0.8$ & 3.4 $\pm$ 0.0 \\ \hline 
  DNC & $J= 1 - 0$ & 76305.727 &  3.66 & 1.6$(-5)$ & 3 & 1.1(5) & CDMS & 0.06 $\pm$ 0.01 & $-$3.0 $\pm 0.8$ & 3.2 $\pm$ 0.1 \\ 
  DNC & $J= 2 - 1$ & 152609.774 &  10.99 & 1.5$(-4)$ & 5 & 1.0(6) & CDMS & 0.18 $\pm$ 0.01 & $-$2.7 $\pm 0.8$ & 3.6 $\pm$ 0.1 \\ 
  HN$^{13}$C & $J=1 - 0$ & 87090.850 &  4.18 & 1.9$(-5)$ & 3 & 1.1(5) & CDMS & 0.07$\pm$ 0.01 & $-$3.0 $\pm 0.8$ & 3.4 $\pm$ 0.1 \\\hline 
  DCN & 
  \begin{tabular}[c]{@{}l@{}} $J=1 - 0$,\\ $F=2-1$\tablefootmark{*}\end{tabular} & 72414.694  & 3.48 & 1.3$(-5)$ & 5 & 3.0(5) & JPL & 0.05 $\pm$ 0.01 & $-$3.1 $\pm 0.8$ & 4.1 $\pm$ 0.1 \\ 
  DCN & $J=3 - 2$ & 217238.538 &  20.85 & 4.6$(-4)$ & 21 & 1.0(7) & CDMS & 0.11 $\pm$0.01 & $-$3.8 $\pm 0.8$ & 4.1 $\pm$ 0.1 \\  
  H$^{13}$CN & 
  \begin{tabular}[c]{@{}l@{}}$J=1- 0$,\\ $F=2-1$\tablefootmark{*}\end{tabular} & 86339.921  & 4.14 & 2.2(-5) & 5 & 3.0(5) & JPL & 0.11 $\pm$0.01 & $-$3.8 $\pm 0.8$ & 4.1 $\pm$ 0.1 \\ \hline 
  \end{tabular}
  \tablefoot{The frequency, upper level energy, $E_{\rm up}/k$, Einstein coefficient, $A_{\rm ij}$, and upper state degeneracy $g_{\rm up}$, are given. The critical density, $n_{\rm crit}$, at $T_{\rm kin} = 20$~K, are also listed \citep{Shirley2015}.
   The rightmost three column give the peak intensity, $T_{\text{MB}}$, central LSR velocity and FWHM of Gaussian fits to a spectrum representing an average of the emission in the 
   covered area in the DR21 region.  
   The values for $A_{\rm ij}$ and $n_{\rm crit}$ are written in the form of $x(-a) = x \times 10^{-a}$. 
   The spectroscopic information is taken, as noted, from the Cologne Database for Molecular Spectroscopy (CDMS) \citep{CDMS} or from the Jet Propulsion Laboratory Spectroscopic catalogue database\citep{JplCat}.
   \tablefoottext{*}{Denotes the spectroscopic properties of the strongest HFS component.}}
  
  \label{tab:Deuterated}
\end{table*}

\subsection{Follow-up observations with IRAM 30m telescope}
To complement the lower $J$-transitions of some of our target molecules
(see Table \ref{tab:Deuterated}), we performed OTF mapping observations using the IRAM 30-m telescope over five nights in 2021 July 15 - July 19 (Project ID: 053-21; PI: I. Barlach Christensen). The OTF maps were carried out with the EMIR E150 and E230 modules in five tiles, each of size of $\sim 120\arcsec \times 120 \arcsec$, outlined by the orange boxes in Fig. \ref{fig:DR21_maps}. 
For each mapping (with two coverages combined in one tile), the scanning velocities were $5\rlap{.}\arcsec5$ s$^{-1}$ and $3\rlap{.}\arcsec8$ s$^{-1}$ for 2-mm and 1.3-mm, respectively. The HPBW is $17\arcsec$ at 145 GHz and $11\arcsec$ at 230 GHz. 

The FTS covered 8 GHz bandwidth in each sideband in both polarization. To obtain a spectral resolution that is comparable to the ground-state transition observations, the FTS backend was used at 195 kHz resolution, which corresponds to 0.38 \kms\ at 154 GHz and 0.25 \kms\ at 231 GHz. 
The spectra are smoothed to a spectral resolution of $\mathrm{d}\upsilon = 0.8$ \kms.
For the E150 observations, the receiver was tuned to 150.25 GHz in the upper inner sideband, to cover the \ce{N2D+} $(2-1)$ and DNC $(2-1)$ transitions. 
The frequency coverage in the 2-mm wavelength band is 132.2--140 GHz in the lower sideband and 148--155.7 GHz in the upper sideband. 
For the E230 observations, the receiver was tuned to 216.75 GHz in the lower outer sideband, to cover the \ce{DCO+} $(3-2)$, DCN $(3-2)$ and \ce{N2D+} $(3-2)$ transitions. The total E230 frequency coverage is 214.4--222.2 GHz for the lower sideband and 230.1--237.9 GHz for the upper sideband. 

The angular resolution of the higher $J$-transition observations of \ce{DCO+}, DCN and DNC (for example $\theta \sim 12\arcsec$ for \ce{DCO+} (3--2), and $\theta \sim 16\arcsec$ for DNC (2--1)) is higher than that of the ground-state transitions (for example, $\theta \sim 34\arcsec$ for \ce{DCO+} (1--0) ). For comparison purposes, the maps of the higher $J$-transitions are convolved to a common beam size, namely that of the ground-state transitions ($\theta = 34\arcsec$). 
However, for the maps shown, the native resolution of the respective lines was retained. 

\subsection{Data Processing}
The data of both the CASCADE and follow-up observations of the DR21 filament were  processed  using the Grenoble Image and Line Data Analysis
Software (\texttt{GILDAS}) package\footnote{\url{http://www.iram.fr/IRAMFR/GILDAS}} \citep{Pety2005,Pety18}. 
Using the Continuum and Line Analysis Single-dish Software (\texttt{CLASS}) program, we regridded the obtained spectra over the mapped areas. 
Using the \texttt{MAP\%LIKE} procedure of \texttt{CLASS}, all images were reprojected to the spatial grid of \ce{DCO+}, that is a resolution of $34\arcsec$.  
For computing the column densities, the scale was converted from $T_{\rm A}^*$ to main beam brightness temperature $T_{\rm MB}$, using a forward efficiency $F_{\rm eff} = 0.95$ and a beam efficiency\footnote{\url{https://web-archives.iram.fr/ARN/feb01/node5.html}} depending on the frequency of the molecule:
\begin{equation}
  B_{\text{eff}} = 1.2\times0.69\times \text{exp}\left(-\left( \frac{4\times\pi\times0.07\nu~[\text{MHz}]}{c~[\text{km s}^{-1} ]} \right)^2\right).
\end{equation}
where $\nu$ is the rest frequency of the line emission and $c$ is the speed of light.
This ensures that each line can be resolved and has the same velocity resolution, as the line widths are typically FWHM $\sim 3 - 4$ \kms\ (see Table \ref{tab:Deuterated}). 

During the observations, we identified a strong narrow spike at 76304.9\,MHz with a frequency shifted by -1~MHz from the emission line of DNC ($\nu = $ 76305.727\,MHz). 
We suspect this spike to be contamination from an automotive long range radar. In order to minimize the contamination caused by this spike, we compared the noise in the artifact free spectral velocity range ($< -30$ \kms\ from the central line) with the contaminated channels ($> 0$ \kms). If the noise level with the artifact increased by 20\%, the spectrum is discarded. 

\subsection{Ancillary data}
Figure~\ref{fig:DR21_maps} presents the dust temperature and \ce{H2} column density \citep{Bonne2023} maps of the DR21 filament first discussed by \citet{Hennemann}, which are based on data taken as part of the Herschel imaging survey of OB Young Stellar objects (HOBYS) program \citep{Motte2010}. 

The CASCADE mapping area is outlined by red dashed lines in both the maps. 
The \ce{H2} column density maps are derived by \cite{Bonne2023} based on convolved dust brightness temperature distributions  at three different wavelengths: 500~$\mu$m, 360~$\mu$m, and 250~$\mu$m. 
The dust temperature is determined from the 160~$\mu$m/250~$\mu$m flux density ratio. The resulting \ce{H2} column density and dust temperature maps are obtained at an angular resolution of approximately $\sim 18 \arcsec$.
For further analysis, both the \ce{H2} column density map and dust temperature maps are convolved to the angular resolution of the present CASCADE study, which is $\theta \sim 34\arcsec$.

\section{Results}
\label{Sec:Results}
\begin{figure*}[htpb!]
  \centering
  \includegraphics[width=\linewidth]{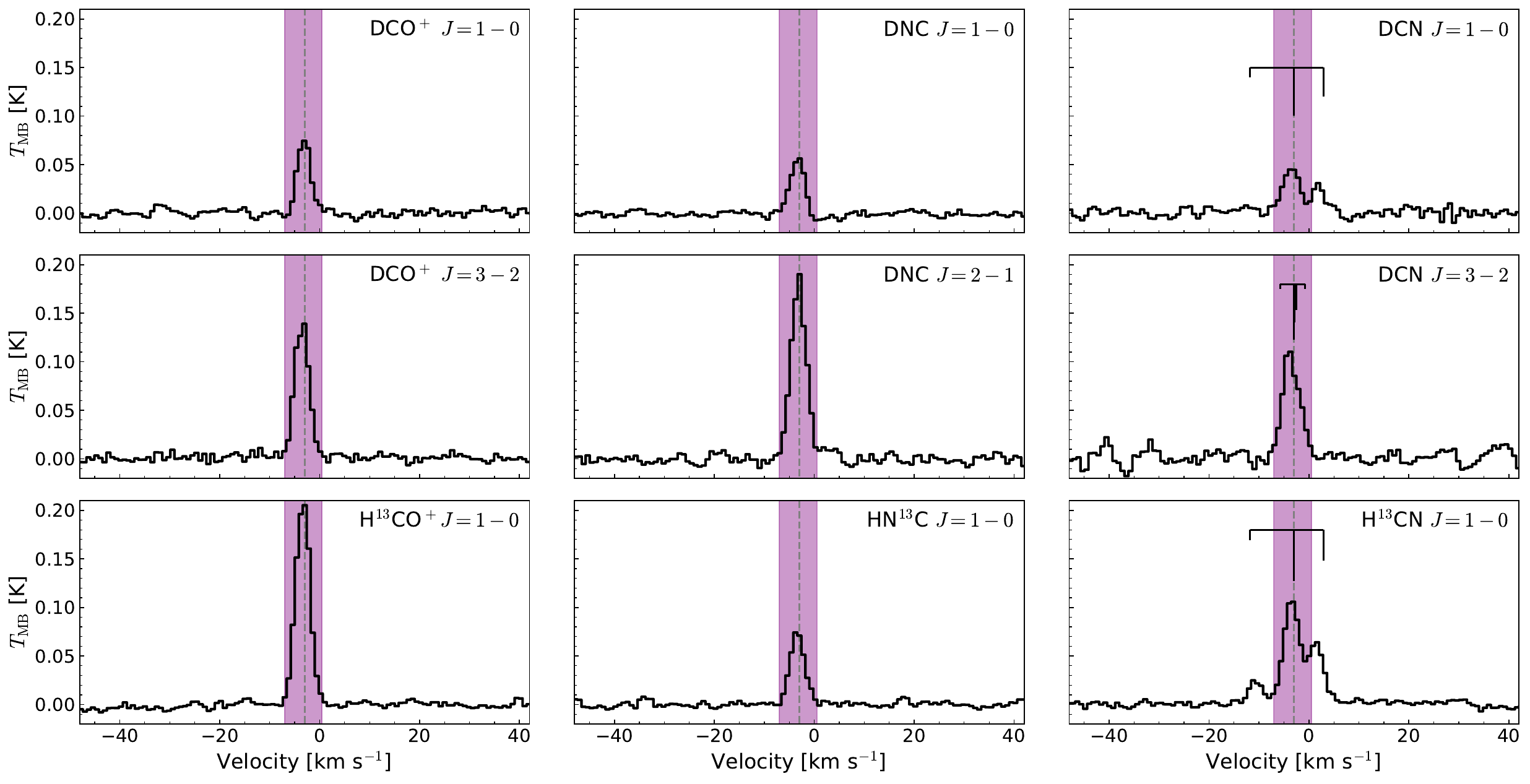}
  \caption{Average spectra of \ce{DCO+}, DNC and DCN, and their \ce{^13C}-isotopologs in the DR21 filament.
  The black line shows the spectra averaged over the entire mapped region of the DR21 filament for different molecules with the purple region showing the velocity range used for  determining integrated intensities. The gray dashed line denotes the main velocity of the filament, $\upsilon_{\rm LSR} = - 3$ \kms. 
  The velocities of the un- or only marginally blended HFS components of \ce{DCN} and \ce{H^13CN} are marked in black, with the relative intensity of the hyperfine components marked by the length. The LSR velocity scale of the DCN and H$^{13}$CN $J = 1-0$ spectra corresponds to the strongest HFS component $F =2-1$.}
  \label{fig:Average_spectra}
\end{figure*}

Using data from the CASCADE survey, we investigate in detail the deuteration along the DR21 filament. 
For this we consider three molecular species that are abundant in the ISM: \ce{HCO+}, HNC, and HCN and their corresponding deuterated species. 
As mentioned in Sect.~\ref{Sec:Introduction}, the three species, and their deuteration, provide insight into distinct physical conditions.
In general, \ce{HCO+}, HNC, and HCN lines are likely to be optically thick in dense molecular clouds, hence to estimate the column densities of these hydrogen-bearing species and to estimate the degree of deuterium fractionation, we utilize lines of their optically thin isotopologs, i.e., \ce{H^13CO+}, \ce{HN^13C} and \ce{H^13CN}. Furthermore, we assume the deuterated isotopologs are optically thin.
Table \ref{tab:Deuterated} summarizes the observed molecules and their transitions. Figure~\ref{fig:Average_spectra} shows spectra of the deuterated species and \ce{^13C} isotopologs averaged over the entire DR21 filament. 
For most of the lines, the hyperfine splitting structure (HFS) was too small and/or not resolved at the resolution of 0.8~\kms, except for the \ce{HCN} (1--0) \ce{^13C} and D-isotopolog lines.  
The line parameters used during the analysis account for the unresolved HFSs.
In the case of the ground-state transition of HCN, \ce{H^13CN} and DCN, the integration is done only focusing on the central and strongest line. 
The emission from this main HFS component is extracted over the same velocity range as that of 
the \ce{DCO+} and DNC lines.
The HFS components of the HCN, \ce{H^13CN} and DCN lines are not considered to ensure that the extracted line emission stems from the $\upsilon_{\rm LSR} = -3$~\kms\ component. 
In contrast, the higher velocity DCN HFS component can be affected by a secondary DR21 velocity component at $\upsilon_{\rm LSR} = 9$~\kms \citep{Dickel1978,Nyman1983}. The resultant line parameters of the deuterated and $^{13}$C isotopologs are presented in Table \ref{tab:Deuterated}. 

\begin{figure*}[htpb!]
  \centering
  \sidecaption
  \includegraphics[width=12cm]{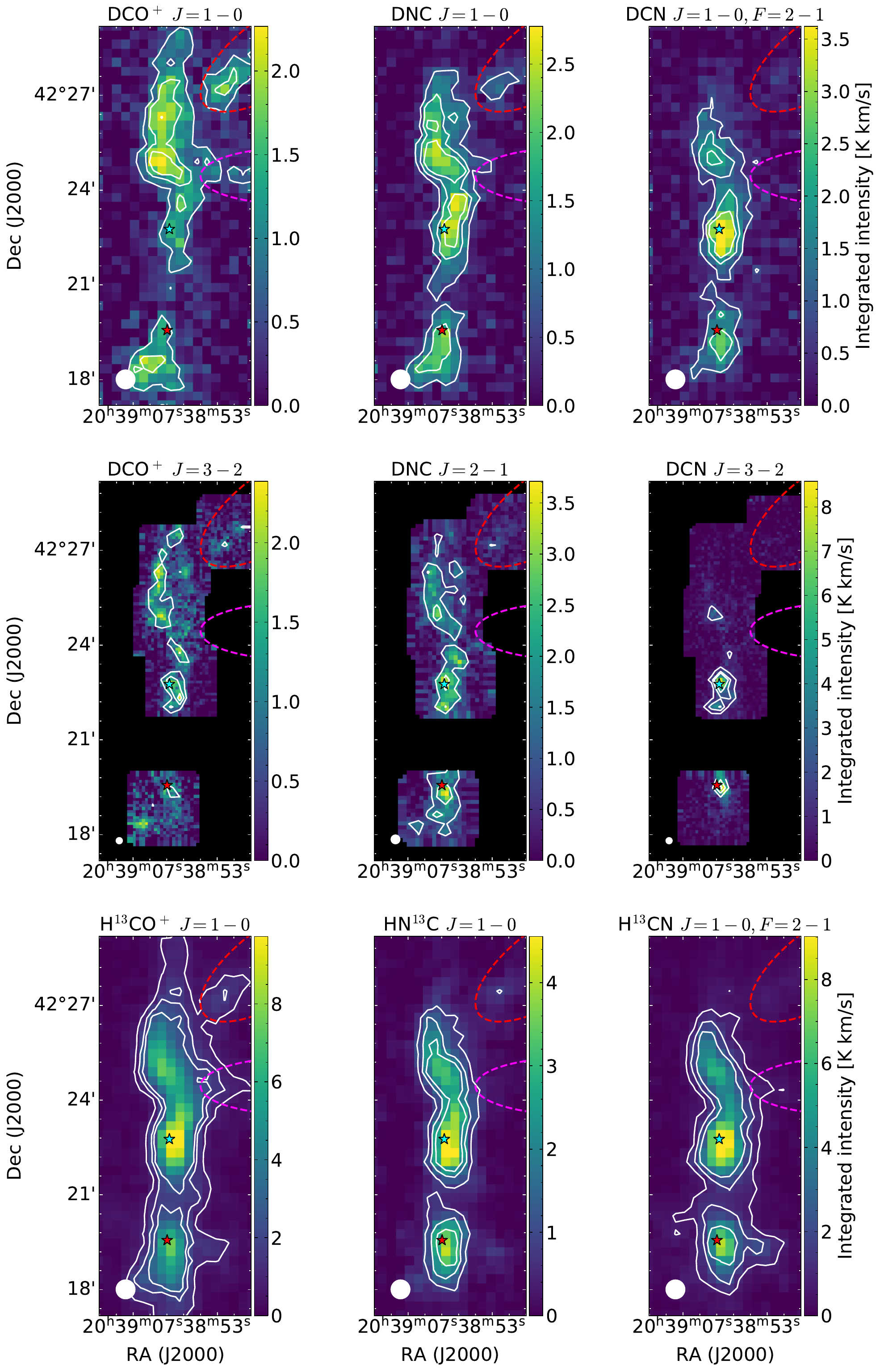}
  \caption{Top and middle rows: The ground-state emission of \ce{DCO+}, DNC and DCN observed with the CASCADE-program and follow-up mapping with the IRAM-30m telescope. Bottom row: Maps for \ce{^13C} isotopologs. The intensities in  the maps are integrated over the velocity range $-7$~\kms\ to $+0.5$ \kms. HPBWs are displayed by the filled white circles in the lower left corner of each panel.
  For example,  the HPBW beam-size  is $34\arcsec$ for the map showing DNC (2--1),  16$\arcsec$ for DCO$^+$ (3--2) and $11$\arcsec for  DCN (3--2).
  The white contours are showing levels of $4\sigma$,  $8\sigma$ and $12\sigma$. The stars mark DR21 Main (red) and DR21(OH) (blue). The locations of the F1 and F3 sub-filament \citep{Schneider10} are indicated by red and magenta dashed curves, respectively. }
  \label{fig:deutmaps}
\end{figure*}

\subsection{Integrated intensity maps of the DR21 filament}
\label{sec:EmissionResults}
To study the spatial distribution of the deuterated molecules in DR21, we created velocity-integrated intensity maps (see Fig. \ref{fig:deutmaps}) over the velocity range from $-7$\,\kms\ to $+0.5$~\kms\ (purple shaded region in Fig. \ref{fig:Average_spectra}). This integration velocity range was chosen around the peak velocity $\upsilon_{\rm peak}\pm 3.5$ \kms, determined from a fit to a spectrum that represents an average of the emission in the 
covered area in the DR21 region (see Table \ref{tab:Deuterated}). As  $\upsilon_{\rm peak}$ has a value $-3$ \kms\ for both  \ce{DCO+} and DNC lines and $-3.5$ \kms\ for the DCN lines, the selected velocity interval from $-7$~\kms\ to $+0.5$ \kms\ ensures that the chosen integration ranges are consistent for all the targeted species. Emission maps of the ground-state transition of the deuterated species and their \ce{^13C} isotopologs are shown in Fig.~\ref{fig:deutmaps}. 
The ground rotational transitions of deuterated molecules (i.e., \ce{DCO+}, DCN, and DNC) show enhanced emission along the DR21 filament. 
For comparison, we also created the intensity maps of the corresponding hydrogenated species shown in Fig.~\ref{fig:CASCADEMainandiso}.
We note that for species like \ce{H^13CN} and \ce{DCN}, that show HFS, the velocity range over which we integrate includes contributions from the other HFS lines as well, see Fig. \ref{fig:Average_spectra}. 
As a result, the column densities subsequently derived for these species from the main HFS component will be overestimated.

\subsubsection{Noise estimation}
For each spectrum per pixel, a baseline is fitted with \texttt{CLASS} by applying a 1st order polynomial fit to an emission-free velocity window in the spectrum near the line of interest.
The emission-free range is determined by visual inspection, aiming for a baseline subtraction in a velocity range of $\sim 20$ \kms. 

We obtain an rms~$\sim 80 - 100$~mK with spectral resolution at 0.8~\kms for the deuterated and the \ce{^13C} isotopologs covered with the CASCADE program.
With the follow-up observations, we obtain an rms~$~\sim 130$~mK with the same spectral resolution for the deuterated species.

The noise in the integrated maps depends on the number of channels over which the molecular line of interest is integrated.
The resulting noise-map for the integrated intensity maps of each target line is calculated as $\sigma_{\rm tot} = \sigma_{\rm channel} \times \sqrt{N_{\rm channels}}$. 
The number of channels, $N_{\rm channels} = \Delta \upsilon/\delta \upsilon$, where $\Delta \upsilon$ is the linewidth and $\delta \upsilon$ is the spectral resolution. 
In the subsequent sections, the analysis discussed are based on the regions with a signal $\geq 4\sigma_{\rm tot}$ for each pixel.

\subsection{Column densities and kinetic gas temperature} 
\label{sec:coldensdet}
Molecular column densities are calculated from the integrated intensities.
The method assumes that the emission is optically thin and also assumes a common excitation temperature for each pixel in all transitions considered.
We assume that all the lines considered are emitted under the conditions of local thermal equilibrium (LTE), which means that $T_{\mathrm{exc}} = T_{\mathrm{rot}} = T_{\mathrm{kin}}$: any line's excitation temperature is equal to a molecular rotation temperature, and both are equal to the kinetic gas temperature.
Under LTE, the column densities can be calculated via the Boltzmann equation \citep{Goldsmith1999}:

\begin{equation}
  {\rm ln}\left(\frac{N_{\rm up}}{g_{\rm up}}\right) = {\rm ln}\left(\frac{N_{\rm tot}}{Q_{\mathrm{\rm rot}}}\right) - \frac{1}{T_{\mathrm{rot}}} \frac{E_{\rm up}}{k_{\rm B}},
  \label{eq:RotDiag}
\end{equation} 
where $N_{\rm up}$ and $g_{\rm up}$ are the population and degeneracy of a transition's
upper energy level, $N_{\rm tot}$ is the total population in all the energy levels, and $Q_{\mathrm{rot}}$ is the rotational partition function of each molecule taken from the CDMS or the JPL spectroscopy database.
$E_{\rm up}$ is the upper level energy, and $k_{\rm B}$ is the Boltzmann constant, and $T_{\mathrm{rot}}$ is the rotation temperature. The parameters used for each molecule are presented in Table~\ref{tab:Deuterated}. We calculated $N_{\rm up}$ for each transition using the integrated intensity maps presented in Fig \ref{fig:deutmaps}. Only those pixels with  emission above $4\sigma$ are considered for the analysis. The upper level population of a transition, $N_{\rm up}$, is calculated using the expression:
\begin{equation}
  N_{\rm up} = \frac{8 \pi k_{\rm B} \nu^2}{h~c^3 A_{\rm ij}} \int T_{\rm mb} \text{d}v.
\end{equation}
Here, $\nu$ is the line's frequency, $h$ is the Planck constant, $c$ is the speed of light and $A_{\rm ij}$ is the Einstein coefficient (see Table \ref{tab:Deuterated}). 
For the high densities of the dense clouds (see Tab.~\ref{tab:Deuterated} for the critical densities of target molecules), the excitation temperature closely follows the kinetic gas temperature so that LTE can be assumed. Assuming a depth of the width of the filament \citep[$d = 0.34$~pc][]{Hennemann}, the column densities, where the deuterated molecules are detected, are at $N(\ce{H2}) \geq 3 \times 10^{23}$~cm$^{-2}$, and the volume densities are $n(\ce{H2}) \geq 3 \times 10^{5}$~cm$^{-3}$.

While Eq.~\eqref{eq:RotDiag} solves for the temperature based on the observation of multiple molecule transitions, not all positions in the map have two detected transitions. 
Therefore, the kinetic gas temperature can be determined using several methods.
\cite{Hacar2020} presented a correlation between the kinetic gas temperature and the integrated intensity ratio of HCN and HNC:
\begin{align}
  T_{\mathrm{kin}}[\text{K}] &= 10 \times \left[ \frac{I(\ce{HCN})}{I(\ce{HNC})} \right] \, \, \text{if} \, \, 1 \leq \left( \frac{I(\ce{HCN})}{I(\ce{HNC})} \right) \leq 4 \label{eq:Tkin_calc1} \\
  T_{\mathrm{kin}}[\text{K}] &= 3 \times \left[ \frac{I(\ce{HCN})}{I(\ce{HNC})} -4 \right] + 40 \, \, \text{if} \, \, \left( \frac{I(\ce{HCN})}{I(\ce{HNC})} \right) > 4
  \label{eq:Tkin_calc}
\end{align}
Due to the different destruction pathways of these isomers, at higher temperatures, HNC is destroyed more effectively, thus increasing the ratio as the temperature increases, in combination with opacity effects. 
The gas kinetic temperature is obtained using this method is used to calculate the column density maps of the deuterated molecules and the main isotopologs.

\begin{figure*}[htpb!]
  \centering
  \sidecaption
  \includegraphics[width=12cm]{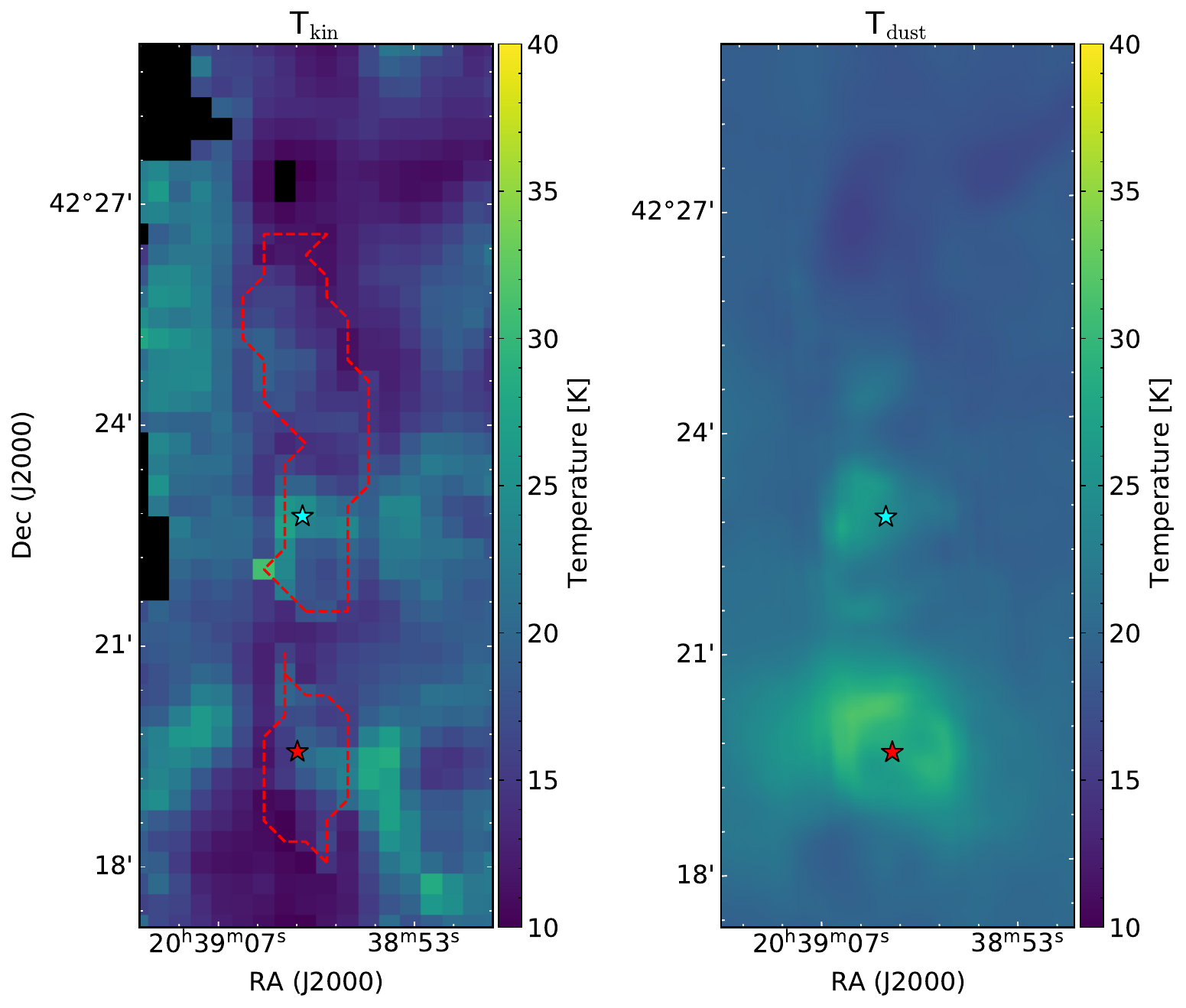}
  \caption{Kinetic gas temperature calculated using the integrated intensity ratio of HCN and HNC for the lower column densities and \ce{H^13CN} and \ce{HN^13C} for the higher column densities (left) and dust temperature in the same region for comparison (right). The red contour shows the column density of $N$(H$_2$) $\geq 1 \times 10^{23}$ cm$^{-2}$. The stars show DR21 Main (red) and DR21(OH) (blue). The right panel shows the comparison of dust temperature in that same region.}
  \label{fig:TempHCNHNC}
\end{figure*}

Using the HCN $(1-0)$ and HNC $(1-0)$ intensity maps (see Fig. \ref{fig:CASCADEMainandiso}) and the above equations, we generated a kinetic temperature map of DR21, shown in Fig.~\ref{fig:TempHCNHNC}. 
For pixels with molecular hydrogen column densities $N$(\ce{H2}) $\geq 1\times 10^{23}$~cm$^{-2}$, HCN and HNC could be optically thick \citep{Beuther2022}. The optical depth of \ce{H^13CO+}, \ce{HN^13C} and \ce{H^13CN} can be seen in App.~\ref{Appendix:OpticalDepth}.
The kinetic temperatures of these pixels were calculated using the intensity ratio of their optically thin isotopologs \ce{H^13CN} and \ce{HN^13C} \citep{Beuther2022,Pazukhin2022}. 
The obtained kinetic temperatures range from 10 to 31 K with a mean value of 17 K (see Fig. \ref{fig:TempHCNHNC}).
Owing to the large beam size of our observations, we do not resolve the complex structures in DR21. 
Hence, the physical properties determined only represent an average value. 
Using Eq.~\eqref{eq:RotDiag}, we generated column density maps of the deuterated molecules as well as their hydrogenated species. 
The column densities are determined pixel-wise using all transitions detected and fitted with the Python package \texttt{scipy.optimize.curve\_fit}. 
The excitation temperature is determined when two transitions are detected, otherwise, the $T_{\rm kin}$ derived from HCN/HNC was adopted. 
The column density maps of \ce{DCO+}, \ce{DNC}, \ce{DCN}, \ce{HCO+}, \ce{HNC} and \ce{HCN} are presented in Fig. \ref{fig:ColDensMap}.
The respective errors are obtained during the fitting, presented in Fig.~\ref{fig:ColDensMap_error}. 
Column densities of the main isotopologs are calculated from their optically thin $^{13}$C isotopologs, assuming that the $^{13}$C/$^{12}$C isotopic ratio is constant within the cloud complex. 
With a Galactocentric distance, $D_{\rm GC}$, of 8~kpc for Cygnus-X, and using the carbon isotope gradient vs. Galactocentric distance relation $^{12}$C/$^{13}$C $= 7.5D_{\rm GC} + 7.6$ \citep{Wilson1994}, we find $^{12}$C/$^{13}$C $= 68$, which we assume in the following analysis.
The gradient recently determined by \citet{Jacob2020} and \citet{Yan2023}, delivers a value of 60, which is similar with the value of 68 adopted for the analysis in this work.
In the analysis that follows we assume no local variation of the isotopic ratio by, for example, isotopic fractionation, selective destruction of \ce{^12C} versus \ce{^13C} or ion-molecule exchange reactions \citep{Wilson1994, Giannetti2014,Jacob2020, Yan2023}.

\subsubsection{Correlations of integrated intensity maps}
The number of pixels above $4\sigma_{\rm tot}$ in the \ce{DCO+}, DNC and DCN of $J=1-0$ emission maps are 148, 115 and 103, respectively. 
Compared with the other species, \ce{DCO+} shows the most extended emission distribution. It is clumpy with several peaks detected toward the northern half of the filament. 
The brightest clump is located $2\rlap{.}\arcmin1$ north of DR21(OH). 
Clumpy \ce{DCO+} emission is also present in the southern region. 
The southern peak is at an offset ${\sim 1\rlap{.}\arcmin2}$ south-east of DR21 Main. 
Apart from the emission along the main filament, an additional bright clump is seen north-west of the main peak and  west of the DR21-filament. 
The emission coincides with the low density, low kinetic gas and dust temperature in-falling gas, named the F1 sub-filament by \cite{Schneider10}. 
Furthermore, \ce{DCO+} (1--0) emission is detected toward the most massive F3 sub-filament (magenta region marked in Fig.~\ref{fig:deutmaps}), which is linked to DR21(OH) \citep{Schneider10}. 
The morphology of the \ce{DCO+} (3--2) emission is similar to that of its ground state transition. However, the peak is $\sim32\arcsec$ south-west of DR21(OH). 

Similar to \ce{DCO+}, the emission of DNC is also filamentary and shows several peaks. The emission is not as extended as the \ce{DCO+} emission toward the north. The F1 sub-filament, marked in red in Figs.~\ref{fig:deutmaps}, is also not as bright in DNC as it is in \ce{DCO+}. The strongest emission is observed from an $\approx 1\arcmin$ sized N--S elongated structure that has  DR21(OH) at its southern end.
A similar sized region is seen in the south of the filament with DR21 Main at its northern end. The map of the DNC (2--1) emission resembles that of the DNC ground-state transition, and the brightest emission also is found close to DR21(OH). 

Conversely, DCN (1--0) shows less extended emission with a strong peak toward DR21(OH). 
Likewise, DCN (3--2) is only detected toward the bright peaks of DCN (1--0). 
Weaker and relatively compact emission is seen surrounding DR21 Main. 
Emission toward the F1 sub-filament is not detected in DCN. 
The DCN (3--2) emission distribution is more compact than that of the (1--0) line and  stronger (factor $\sim 2$).
Its two peaks are found toward DR21(OH) and DR21 Main.

\begin{figure}[htpb!]
    \centering
    \includegraphics[width=\linewidth]{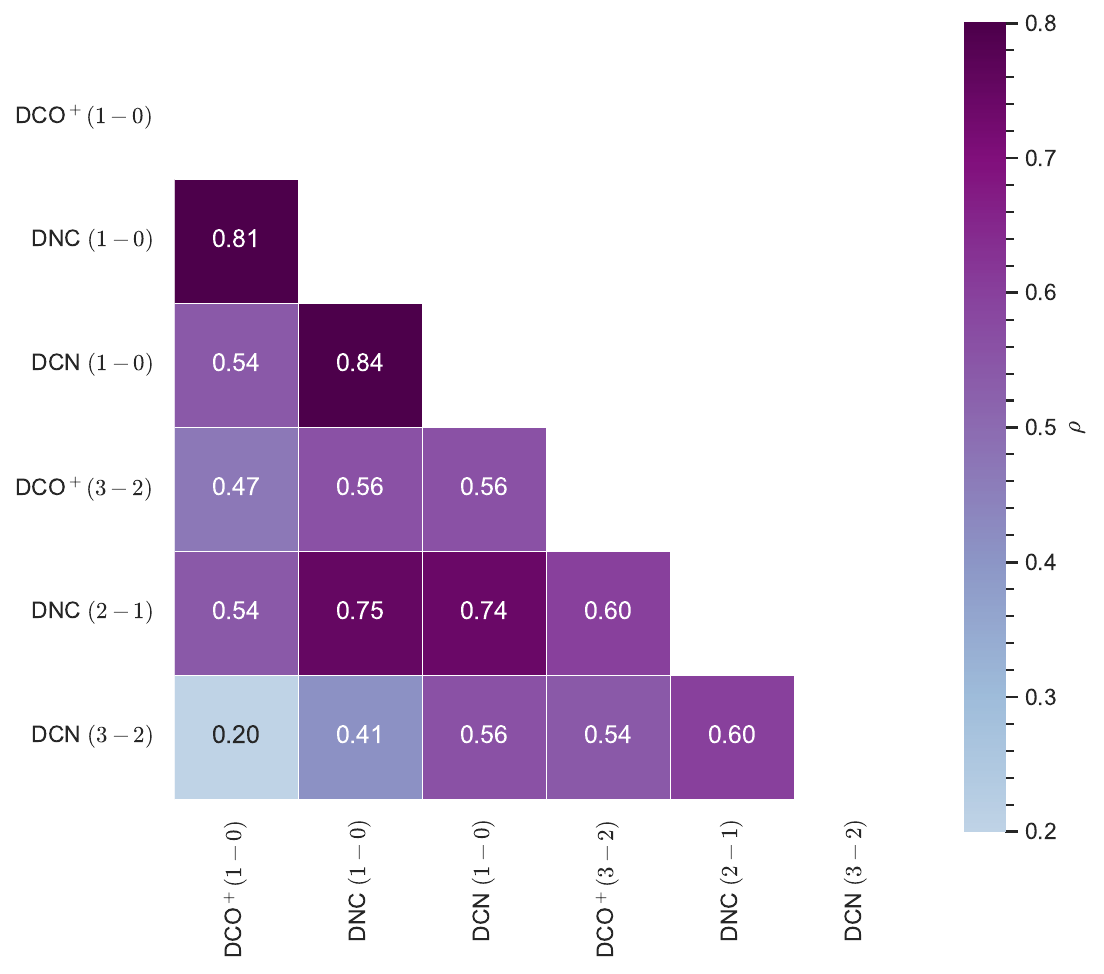}
    \caption{Cross-correlations of integrated intensity maps of \ce{DCO+} (1--0), DNC (1--0), DCN (1--0), \ce{DCO+} (3--2), DNC (2--1) and DCN (3--2).}
    \label{fig:Cross_corr_fig}
\end{figure}

To quantify the morphological similarities between \ce{DCO+}, \ce{DNC} and \ce{DCN}. The cross-correlation, $\rho$, of the distributions of the integrated intensities of pairs of molecules are calculated, pixel-by-pixel, as \citep{Guzman2018,Li2022}:
\begin{equation}
    \rho_{12} = \frac{\sum_{i,j} I_{1,ij} I_{2,ij} w_{ij} }{\left( \sum_{i,j} I_{1,ij}^2 w_{ij} \sum_{i,j} I_{2,ij}^2 w_{ij} \right)^{1/2}}\,.
\end{equation}
Here, $I_{1,ij}$ and $I_{2,ij}$ are the integrated intensities of molecules 1 and 2 in pixel $(i,j)$. 
$w_{ij}$ is the weight of the line emission detection. If either emission from molecule 1 or 2 is detected above $4\sigma_{\rm tot}$ in pixel $(i,j)$, $w_{ij} = 1$. 
If both molecules not detected in pixel $(i,j)$, $w_{ij} = 0$ and not accounted for in the cross-correlation. 
The cross-correlation values are summarized in Fig.~\ref{fig:Cross_corr_fig}. 
This analysis confirms the result of visual inspection of Fig.~\ref{fig:deutmaps}, that the morphology of DNC can be considered as "intermediary" between those of \ce{DCO+} and \ce{DCN}. For example, the cross-correlation of \ce{DCO+} (1--0) and DCN (1--0) is 0.54 whereas DNC (1--0) shows a stronger correlation with \ce{DCO+} (1--0) and \ce{DCN} (1--0)  of $\rho = 0.81$ and 0.84, respectively.

\begin{figure}[h!]
  \centering
  \includegraphics[width=\linewidth]{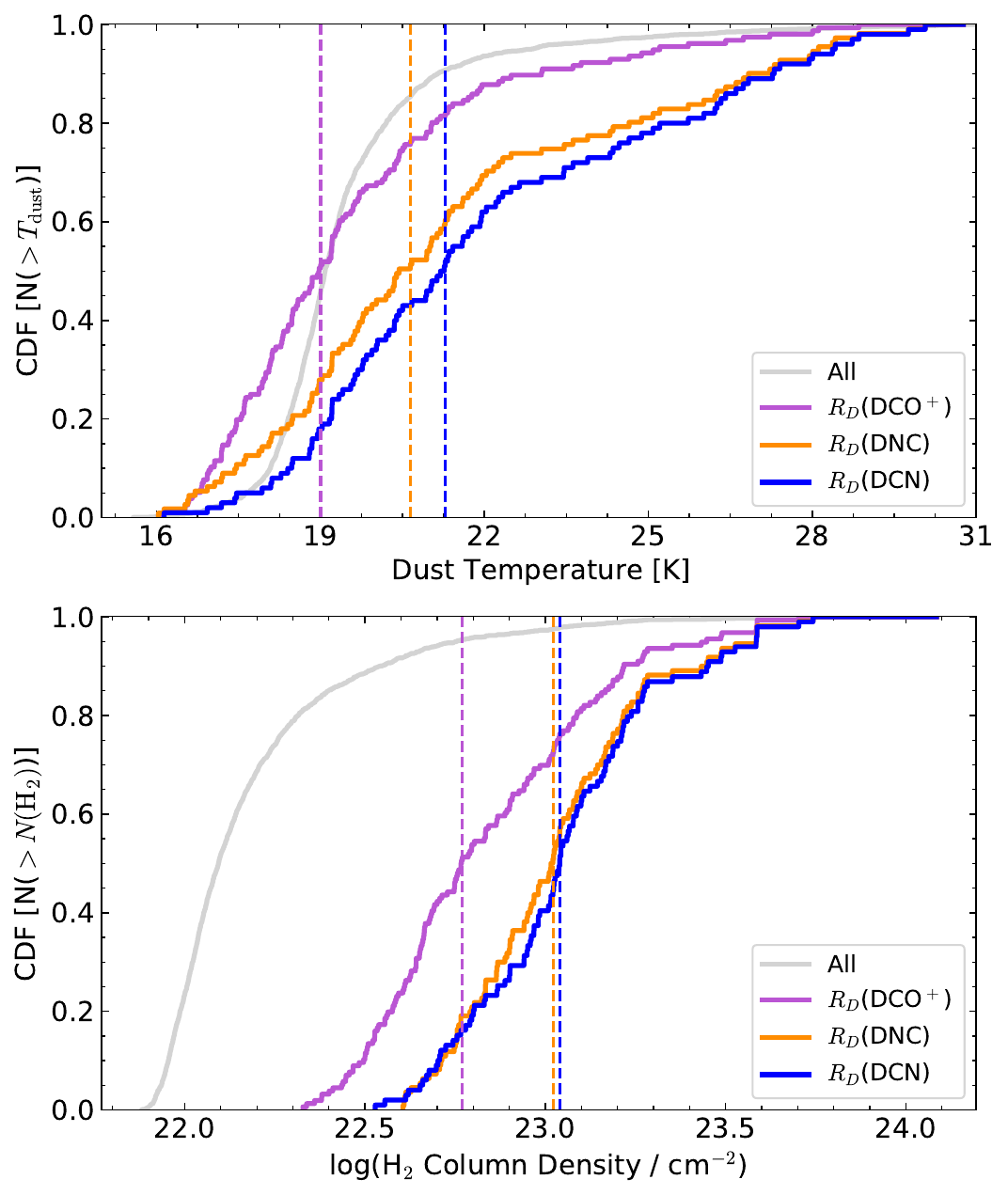}
  \caption{Cumulative distribution function (CDF) of dust temperature and \ce{H2} column density where \ce{DCO+} (purple), DNC (orange) or DCN (blue) is detected in the DR21 filament. 
  The dust temperatures and \ce{H2} column density maps are taken from \citet{Bonne2023}.
  The horizontal dashed lines marks the mean dust temperature or column density of the sampled distributions, colorized with the respective detected molecular ratio.}
  \label{fig:phys_distribution}
\end{figure}

\ce{DCO+} (1--0) seems to originate from gas with different physical conditions than DNC (1--0) and DCN (1--0). 
Figure~\ref{fig:phys_distribution} shows the cumulative distributions of dust temperature and \ce{H2} column densities sampled in the pixels where $R_D$(\ce{DCO+}) (purple), $R_D$(DNC) (orange) and $R_D$(DCN) is detected ($\geq4\sigma_{\rm tot}$). 
A simple Kolmogorov-Smirnov (KS)-test indicates a similar distribution for \ce{DNC} (1--0) and DCN (1--0) ($p < 0.05$). 
The mean dust temperature, where \ce{DCO+} (1--0) is detected, is $T_{\mathrm{dust}} = 19$ K and $N$(\ce{H2})$ = 5.9 \times 10^{22}$ cm$^{-2}$. 
The dust temperature and column density in regions in which DNC (1--0) or DCN (1--0) are detected, is higher. For DNC (1--0), $T_{\mathrm{dust}} = 20.6$ K and $N$(\ce{H2})$ = 1.0 \times 10^{23}$ cm$^{-2}$ and for DCN (1--0) $T_{\mathrm{dust}} = 21.3$ K and $N$(\ce{H2})$ = 1.1 \times 10^{23}$ cm$^{-2}$. The mean values are marked in Fig.~\ref{fig:phys_distribution} with the respective color of the deuterated molecule.

Emission lines from all three deuterated molecules are observed at positions with column densities of $N$(H$_2$) $> 3 \times 10^{22}$ cm$^{-2}$. The \ce{DCO+} (1--0) and DNC (1--0) integrated maps show extended emission features spatially coinciding well with that of the filamentary structure seen in the \ce{H2} column density map. 

The morphological similarities and differences observed between the three molecules suggest local variations in the formation and excitation of their transitions. 
These variations could be influenced by the local density and temperatures of the emitting gas.
The dependencies of each molecule with the column density and the dust temperatures are analyzed further in Sect. \ref{sec:AnalysisMaps}.

\section{Analysis}
\label{Sec:Analysis}
In the following sections, we investigate deuteration across the DR21 filament.

\subsection{Fractional abundances of \texorpdfstring{DCO$^+$}{DCO+}, DNC and DCN}
The fractional abundance is defined as $X$(XD)$= N$(XD)/$N$(\ce{H2}) where $N$(XD) is the column density of a deuterated  molecular species. 
The molecular hydrogen column densities, $N$(\ce{H2}), are adopted from the high-resolution HOBYS column densities described in Sect.~\ref{Sec:Results} and convolved to an angular resolution of $34\arcsec$.
In order to estimate the fraction of deuteration in DR21, we use pixel-by-pixel column density estimates, obtained as described in Sec. \ref{sec:coldensdet}, of the deuterated species presented in Table~\ref{tab:Deuterated}.

\begin{figure*}[htbp!]
  \centering
  \includegraphics[width=\linewidth]{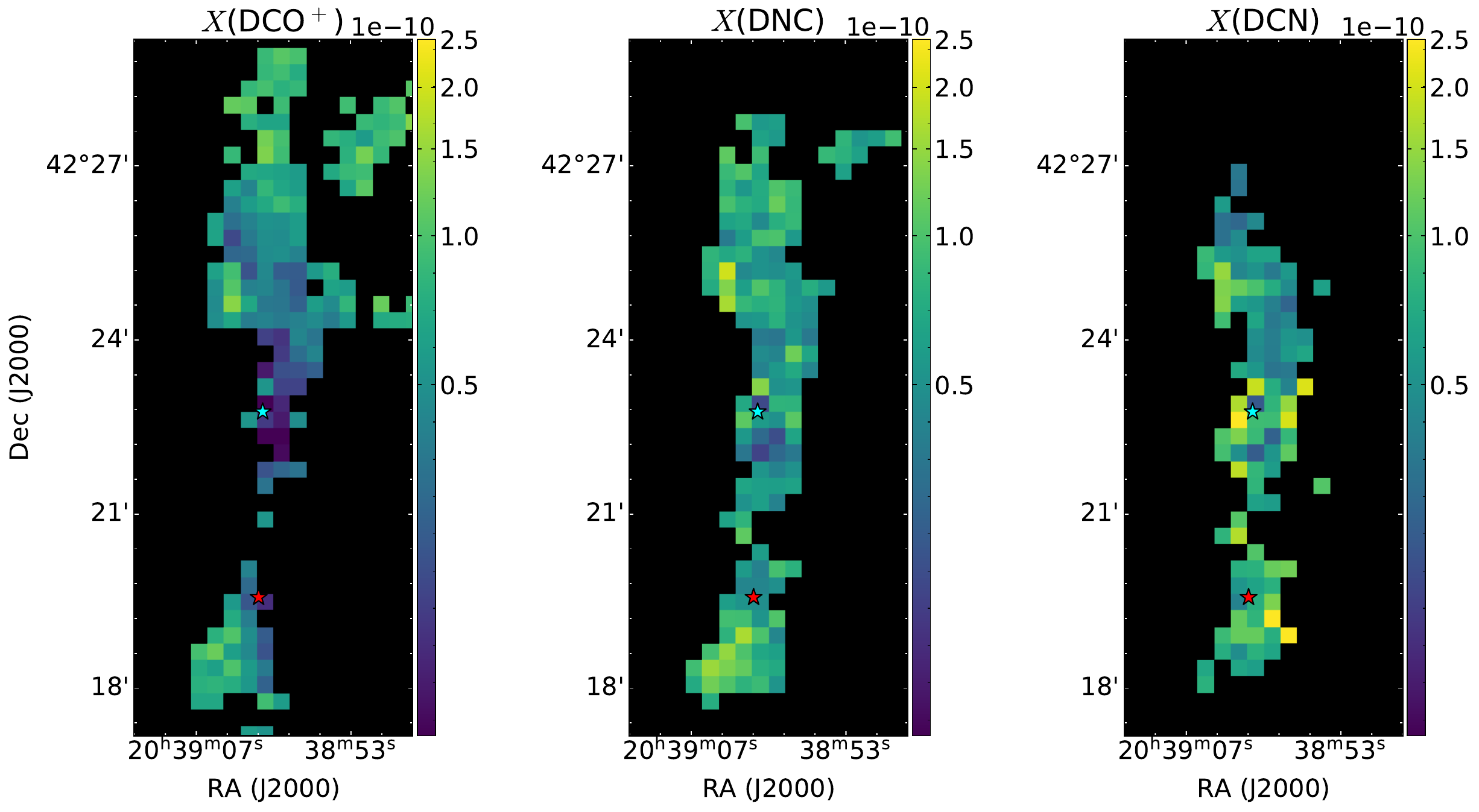}
  \caption{Abundance maps of $X$(\ce{DCO+}), $X$(DNC) and $X$(DCN). The stars mark the positions of DR21 Main (red) and DR21(OH) (blue).}
\label{fig:abundance_maps}
\end{figure*}

Figure~\ref{fig:abundance_maps} shows the abundance maps of $X$(\ce{DCO+}), $X$(DNC) and $X$(DCN).
The abundances of these species are similar, with median values of $X$(\ce{DCO+}) $= 5.9_{1.4}^{10} \times 10^{-11}$, $X$(DNC) $= 6.5_{3.2}^{9.8} \times 10^{-11}$ and $X$(DCN) $= 6.8_{1.9}^{12} \times 10^{-11}$.
Though the median abundances of these three molecular species are indistinguishable within the interquartile range, their spatial distributions are different. 
\ce{DCO+} shows reduced abundances toward DR21(OH) and DR21 Main. $X(\ce{DCO+})$ increases toward the north region of DR21 and south-east region of DR21 Main.
On the contrary, $X$(DCN) exhibits a minor rise toward DR21(OH), while maintaining a relatively constant abundance throughout the DR21 filament.

\begin{table}
  \centering
    \caption{Abundances of \ce{DCO+}, DNC and DCN toward DR21, TMC-1, L134N, IRAS 16293$-$2422 (LM) and G210.82$-$19.47N (G210).} 
    \begin{tabular}{p{1.1cm}p{1.1cm}p{1.1cm}p{1.1cm}p{1.1cm}p{1.1cm}} \hline \hline \\ [-0.1mm] 
     & DR21 & TMC-1                 & \multicolumn{1}{c}{L134N}                 &  LM  & G210      \\ \hline 
    $X(\ce{DCO+})$ &  $5.9$    & $4.4^{a}$ & $9.0^{a}$ &  $1.5^{b}$ & $0.37^{c}$ \\ [0.2cm] 
    $X(\ce{DNC})$  &  $6.5$    & $71^{a}$ & $41^{a}$ &  $0.5^{b}$ & $0.13^{c}$ \\ [0.2cm] 
    $X(\ce{DCN})$  &  $6.8$    & $3.7^{a}$ &  $27^{a}$ & $2.5^{b}$ & -   \\ [0.2cm]\hline 
    \end{tabular}
  \tablefoot{The abundances are written in the form of $x$, where = $x \times 10^{-11}$.}
   \tablebib{$^{a}$\cite{Turner2001} for TMC-1 and L134N, $^{b}$\cite{vanDishoeck1995} for IRAS 16293-2422, $^{c}$ \cite{Tatematsu2020} for G210.82-19.47N.} 
  \label{tab:AbundanceComp}
\end{table}

Table \ref{tab:AbundanceComp} summarizes the abundances of deuterated species in DR21 compared with values determined for various other sources. This includes the extensively studied Taurus Molecular Cloud 1 (TMC-1) which is a quiescent dark cloud containing starless cores \citep{Pratap1997} as well as another dark cloud, L134N \citep{Dickens2000L134N}, shares similar temperature ($T_{\mathrm{dust}} < 15$~K, \citealt{Gratier2016}) and density ($n(\ce{H2}) \sim 10^{6}~\ce{cm}^{-3}$, \citealt{Turner2001}) characteristics to the DR21 filament. However, the primary distinction between these dark clouds lies in their chemical composition: the TMC-1 position is that of the clouds' cyanopolyyne peak, whose chemistry is enriched in carbon-chain molecules, while L134N is abundant in smaller molecules and oxygen-bearing species \citep[][and references therein]{Dickens2000L134N}. 
G210.82--19.47N is a solar-mass star-forming core \citep{Tatematsu2020} and IRAS 16293$-$2422 is a low mass class 0 binary protostar with a rich molecular inventory surrounding a hot corino \citep{Girart, Murillo,Kahle2023}.

Compared to DR21, the abundances of most of the deuterated species studied here, are significantly higher in the dark clouds, but also in the translucent clouds. 
Both dark clouds have an $X(\ce{DCO+})$ that is comparable to that of DR21, but higher $X(\ce{DNC})$,
whereas $X(\ce{DCN})$ is comparable in DR21 and TMC-1, but much lower than in L134N. 
The generally higher degrees of deuteration observed in TMC-1 and L134N relative to DR21 is a consequence of the lower gas temperatures ($T_{\mathrm{kin}} = 10$~K) that characterize these regions \citep{Turner2001}, which, as discussed in Sect.~\ref{Sec:Introduction}, facilitates the deuterium enrichment.

In the low-mass star-forming regions IRAS 16293$-$2422 and G210.82$-$19.47N, the abundances are approximately an order of magnitude lower than the abundances measured in the DR21 filament.
It is important to point out that accurate abundances rely on a reliable reference for $N$(\ce{H2}). 
The challenge of comparing abundances is compounded by the absence of a singular approach to determining column densities of \ce{H2}.

To summarize: we find similar abundances (indistinguishable within the interquartile range) only for \ce{DCO+} in the DR21 filament as for the dark clouds and even for IRAS 16293$-$2422. 
Both TMC-1 and L134N have lower dust temperatures compared to the DR21 filament. The molecular cloud TMC-1 lacks, contrary to DR21, newly formed OB stars and also has a high extinction, resulting in an absence of a UV radiation inside of it.
Our findings for the DR21 filament indicate that, in terms of the abundances of deuterated molecules, it comprises an intermediate stage between the low-mass star-forming regions and the large-scale structure of dark clouds.

\begin{figure*}
    \centering
    \includegraphics[width=\linewidth]{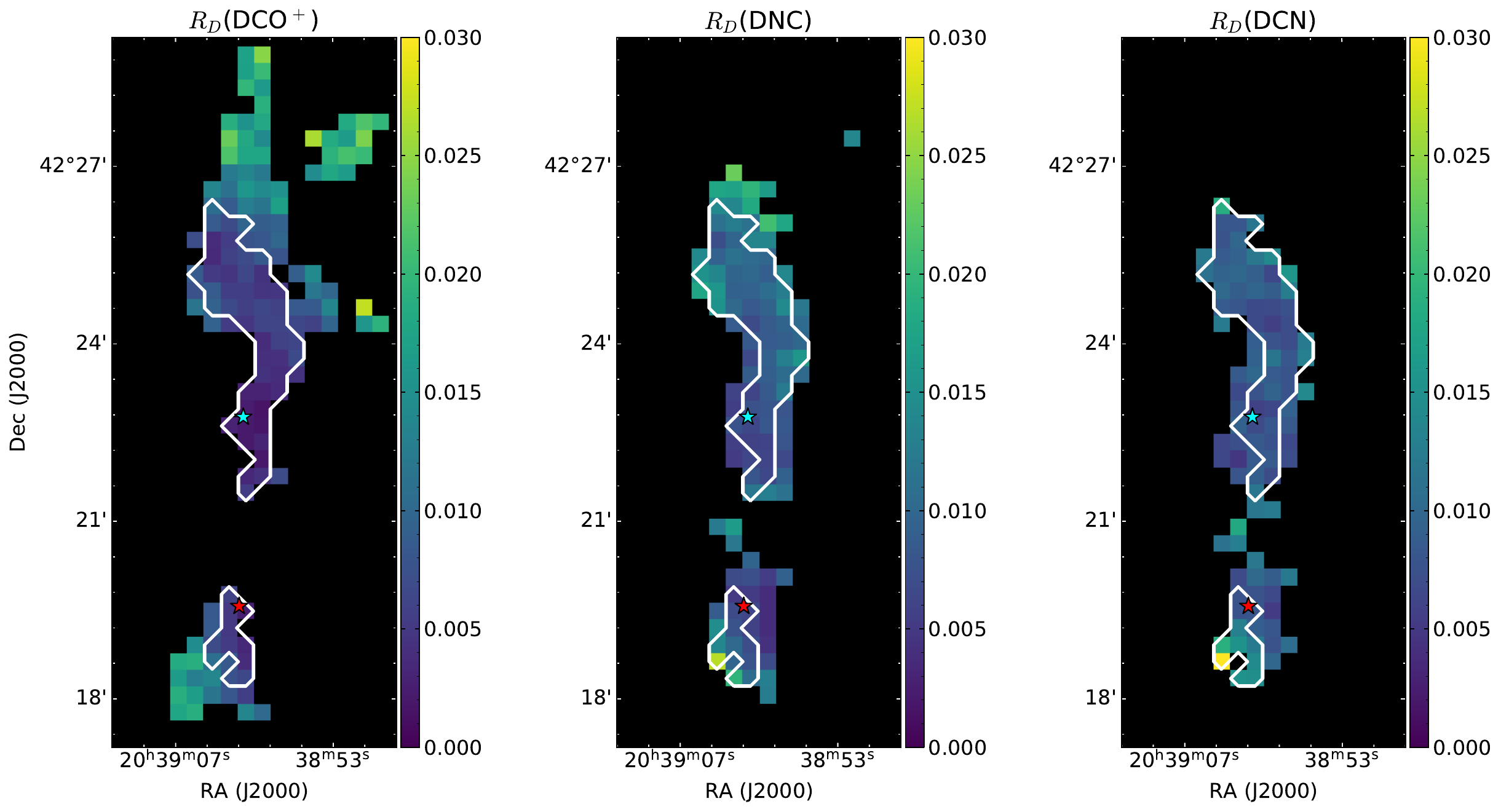}
    \caption{Left to right: $R_D$(\ce{DCO+}), $R_D$(\ce{DNC}) and $R_D$(\ce{DCN}) in the DR21 filament. The stars show DR21 Main (red) and DR21(OH) (blue). The white contours show the pixels, for which $R_{D}$ could be determined for all three species.}
    \label{fig:DHmaps}
\end{figure*}
\subsection{Deuteration degree in the DR21 filament}
\label{sec:AnalysisMaps}
In this section, we estimate the degree of deuteration (or simply deuteration) in the DR21 filament. Deuteration is defined as $R_D$(D)$\,=N$(XD)/$N$(XH), where $N$(XD) and $N$(XH) are the  column densities of the D-bearing isotopolog and the H-bearing (main) isotopolog, respectively. 
The column density maps and the respective relative error maps are shown in Figs. \ref{fig:ColDensMap} and \ref{fig:ColDensMap_error}. 
The distributions of $R_D$(\ce{DCO+}), $R_D$(\ce{DNC}) and $R_D$(\ce{DCN}) are presented in Fig.~\ref{fig:DHmaps}. For further analysis, the pixel-by-pixel $R_D$ values obtained from these maps are used. 

The deuteration in all three species shows smooth distributions, with average values of $0.012 \pm 0.003$, $0.014 \pm 0.003$ and $=0.010 \pm 0.002$, for $R_D$(\ce{DCO+}), $R_D$(DNC), and $R_D$(DCN), respectively.
Thus the overall degree of deuteration in the DR21 filament is $\sim 1$\%.
Similarly, \cite{Yang2023} finds comparable deuteration between $R_D$(\ce{DCO+}), $R_D$(DNC), and $R_D$(DCN) toward the deuterated peaks, despite different detection rates.
While significant amounts of $R_D$(\ce{DCO+}) are measured across the entire filament spanning $N(\ce{H2})$ between $2.1\times 10^{22}$~cm$^{-2}$ and $1.2\times 10^{24}$~cm$^{-2}$, deuteration is particularly enriched in regions of lower \ce{H2} column densities. This is consistent with the increase in $X$(\ce{DCO+}) toward dark clouds and translucent clouds.

The median \ce{H2} column density for which \ce{DCO+} is detected, is $N$(\ce{H2}) $= 5.9 \times 10^{22}$~cm$^{-2}$, whereas DNC and DCN are detected at higher \ce{H2} column densities ($N$(\ce{H2}) $= 1.1 \times 10^{23}$~cm$^{-2}$). 
This further illustrates that \ce{DCO+} samples regions that are less dense in comparison to that probed by DCN and DNC (see also Fig.~\ref{fig:phys_distribution}), as a consequence of the lower critical densities (see Tab.~\ref{tab:Deuterated}).
\cite{Taniguchi2024} found that at lower $N(\ce{DCN})$/$N(\ce{DCO+})$ ($\sim 4-7$, comparable to what we find in DR21, see Fig.~\ref{fig:DDwithTempandColD}) favor models with higher densities.

\begin{table}
  \centering
  \caption{The mean deuteration fraction of each molecule and the range of deuteration in the DR21-filament (where detected $\geq 4 \sigma_{\rm tot}$ level) and the overlapping region, where all three $J = 1-0$-transitions of the deuterated molecules are observed.}
  \begin{tabular}{lcc|c} \hline \hline \\ [-0.1mm]
       & \multicolumn{2}{c|}{Entire DR21} & Overlapping \\
     D/H & Mean & Range & Mean \\ \hline 
     $R_D$(\ce{DCO+}) & 0.012 $\pm$ 0.003 & 0.002  -- 0.043 & $0.006 \pm 0.001$ \\ [0.2cm] 
     $R_D$(DNC) & 0.014 $\pm$ 0.003 & 0.006 -- 0.056 & $0.014 \pm 0.003$\\ [0.2cm]
     $R_D$(DCN) & 0.010 $\pm$ 0.002 & 0.005 -- 0.031 & $0.010 \pm 0.002$\\ [0.2cm]
     \hline
  \end{tabular}
  \label{tab:DHratiodependence}
\end{table}

In the overlapping region (see Fig. \ref{fig:DHmaps}), where all three deuterated molecules are detected, the highest deuteration is seen for DNC with $R_D$(DNC) $= 0.014 \pm 0.003$, which is comparable to $R_D$(DCN)$ = 0.010 \pm 0.002$. 
In contrast, $R_D$(\ce{DCO+}) is a factor $\sim 2$ lower ($ = 0.006 \pm 0.001$) in this region. 
For a summary, see Table~\ref{tab:DHratiodependence}.
The overlapping region, in which emission from all three molecules is detected has elevated  column densities. 
Neither $R_D$(DNC) nor $R_D$(DCN) are detected in the region that shows the highest $R_D$(\ce{DCO+}).
This region coincides with lower dust temperatures and higher \ce{H2} column densities (see the northern region in Fig.~\ref{fig:DR21_maps} for reference).

\begin{figure*}[htbp!]
  \centering
  \includegraphics[width=\linewidth]{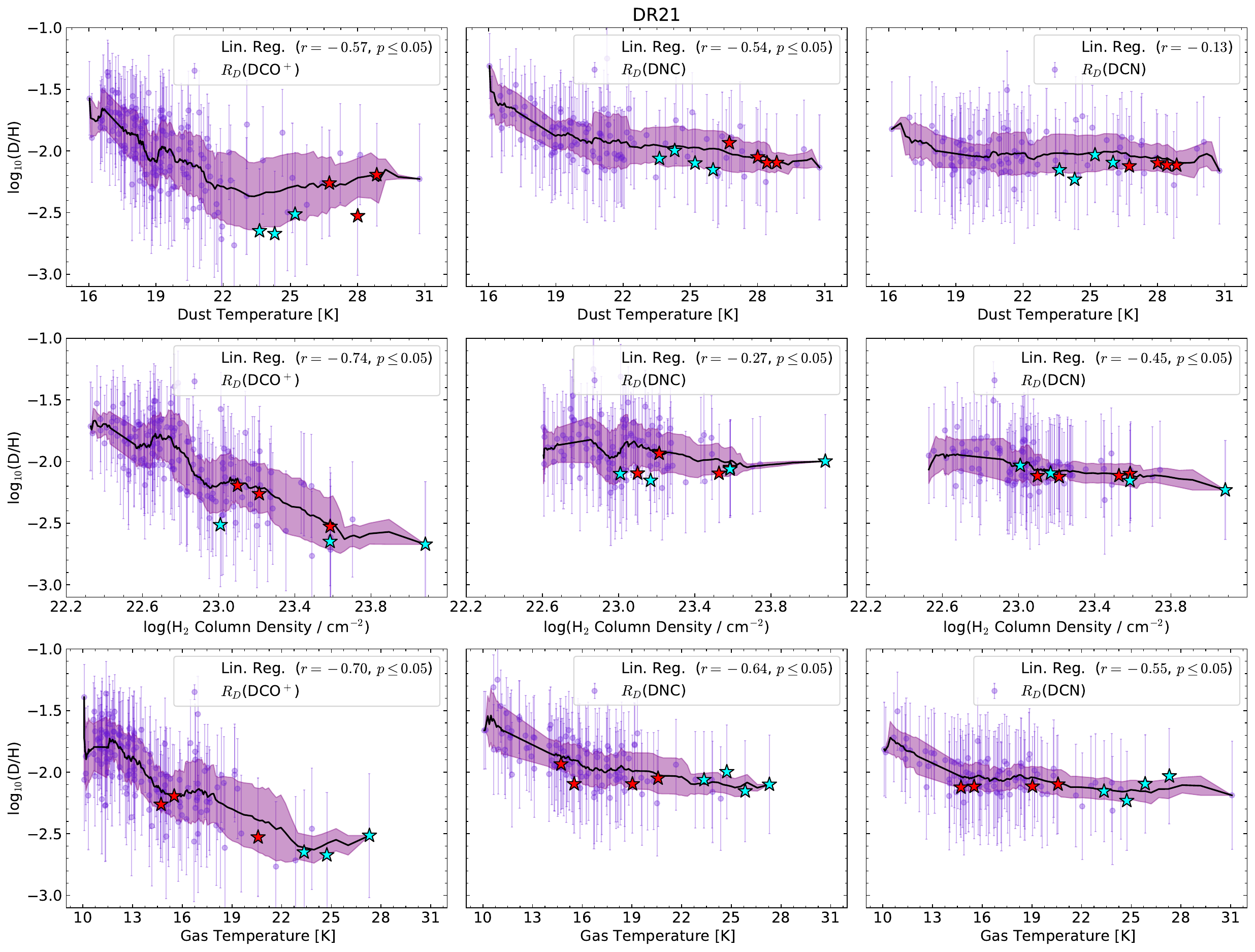}
  \caption{Degrees of deuteration $R_D$(\ce{DCO+}) (left column), $R_D$(DNC) (middle column) and $R_D$(DCN) (right column) with dust temperature (top row) and \ce{H2} column densities (middle row) \citep{Bonne2023} and kinetic gas temperature (bottom row) in the DR21 filament. Each point corresponds to the pixel-wise deuteration derived from the $J=1-0$ and $J=2-1$ or $J = 3-2$ transitions of \ce{DCO+}, DNC and DCN where each of the species were detected at an rms noise level $\geq4\sigma_{\rm tot}$. The black line shows the running mean, while the  purple ranges marks the standard deviation.
  The red and blue stars mark the surroundings of DR21 Main and DR21(OH), respectively.
  }
  \label{fig:Beautiful_corr}
\end{figure*}

To understand the differences in morphology, we compare the spatial variations in deuteration with the distributions of dust temperatures, molecular hydrogen column densities and kinetic gas temperatures along the DR21 filament.
The resulting comparisons are presented in Fig.~\ref{fig:Beautiful_corr}.
The surroundings of DR21 Main and DR21(OH) are marked in their respective colors from Fig.~\ref{fig:deutmaps}. 
The surroundings of DR21 Main and DR21(OH) in this work are $34\arcsec \times 34\arcsec$, corresponding to 4 pixels, based on the spatial resolution comparable to the surroundings examined by \cite{Koley2021}. 
The molecular D/H-ratios show clear differences in their correlations with the dust temperature and \ce{H2} column density. 

\begin{figure}[htpb!]
    \centering
    \includegraphics[width=\linewidth]{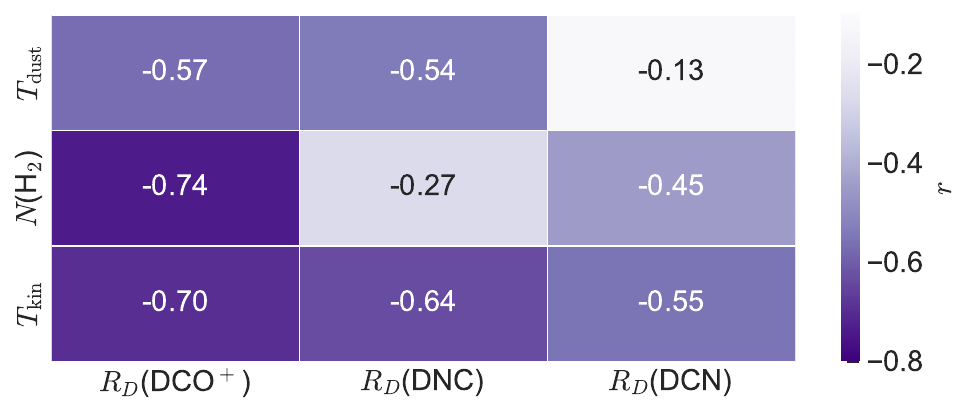}
    \caption{Overview of the Pearson correlation coefficients ($r$) estimated for cross-correlations between \ce{DCO+}, DNC and DCN, and $T_{\rm dust}$, $N(\ce{H2})$ and $T_{\rm kin}$.}
    \label{fig:Cross_corr_ratio}
\end{figure}

Figure~\ref{fig:Cross_corr_ratio} presents an overview of the Pearson correlation coefficients between $R_{D}$(\ce{DCO+}), $R_{D}$(DNC) and $R_{D}$(DCN), and $T_{\rm dust}$, $N(\ce{H2})$ and $T_{\rm kin}$.
We note the strongest anti-correlations ($|r| \geq 0.70$) between $R_D(\ce{DCO+})$ with \ce{H2} column density and kinetic gas temperature. 
At $T_{\mathrm{dust}}\leq 22$~K, the trend for $R_D$(\ce{DCO+}) strongly decreases, whereas at $T_{\mathrm{dust}}\geq 22$~K, the dependence of $R_D$(\ce{DCO+}) on $T_{\mathrm{dust}}$ flattens.
Similar to the dependence on $T_{\mathrm{dust}}$, $R_D$(\ce{DCO+}) shows a decreasing trend at $T_{\rm kin} \leq 22$~K, which flattens out beyond $T_{\mathrm{kin}} = 22$~K. However, the limited number of data points present in this regime prevents us from making definitive conclusions about this correlation.
Both $R_D$(DNC) and $R_D$(DCN) display moderate to weak anti-correlations with $T_{\rm dust}$ and $N$(\ce{H2}) and $T_{\rm kin}$, where the trends with $R_D(\ce{DCN})$ are particularly flatter (see the right-hand panel Fig.~\ref{fig:Beautiful_corr}). 
This is unsurprising because the DCN emission is detected toward a more compact region and hence shows a smaller range of deuteration.

\cite{Gerner2015} conducted a similar study on deuteration in these species for 59 high mass star-forming regions. These authors determined that luminosity (as a proxy for temperature) and \ce{H2} column density did not show a significant correlation with any of the three ratios.
On average \cite{Gerner2015} reported a deuteration of 0.0025 for $R_D$(\ce{DCO+}), 0.005 for $R_D$(DNC) and 0.02 for $R_D$(DCN). These estimates are a factor of $\sim 4$ and $\sim 2$ lower than our estimates of $R_D$(\ce{DCO+}) and $R_D$(DNC) while that for $R_D$(DCN) is a factor $\sim 2$ higher than our estimates in DR21.
Unlike the high mass star-forming regions studied by \cite{Gerner2015}, the DR21 filament shows a stronger anti-correlation with \ce{H2} column density and $R_D$(\ce{DCO+}).
However, we note that there are key differences between both studies; we compute $R_D$ using mapping observations whereas \cite{Gerner2015} performed pointed observations toward their sample of star forming regions. 
Hence, the D/H-ratios derived by these authors probe only the average properties of the targets studied.
Moreover, these authors computed the column densities of these deuterated species using fixed values for the $T_{\rm exc} (=T_{\rm rot})$ based on their evolutionary stages\footnote{$T=15$~K for IRDCs, $T=50$~K for HMPOs, $T=100$~K for HMCs and $T=100$~K for UC\hii.}.
Among these evolutionary stages, the IRDCs have temperatures most similar to DR21 among these evolutionary stages, $T_{\mathrm{kin}} = 15$~K.
The typical values for $R_D(\ce{DCO+})$, $R_D(\ce{DNC})$ and $R_D(\ce{DCN})$ derived by \citet{Gerner2015} for the subset of sources classified as IRDCs in their sample are 0.007, 0.008 and 0.015, respectively. 
These values are comparable to the mean values of the DR21 filament within the errors. 
\cite{Hennemann} investigated the evolutionary stages along the DR21 filament, where they found an increase in evolutionary gradient from the northern region of DR21 toward the DR21(OH). 
The deuteration fraction along the filament follows a trend of higher deuteration toward the younger region, but a detailed analysis will be done in a forth-coming paper.

\section{Discussion}
\label{Sec:Discussion}
The analysis presented thus far reveals that the strongest dependencies found are between the different physical parameters and $R_D(\ce{DCO+})$, while much weaker correlations were found for $R_D(\ce{DNC})$ and $R_D(\ce{DCN})$. 
Therefore, in the following sections, we explore the impact of the physical conditions, if any, on the chemistry of these species.

\subsection{Impact of physical conditions on \texorpdfstring{$R_D$(\ce{DCO+})}{RDCOp}}
\label{sec:DCOHCOcolumndensity}
\begin{figure*}[htbp!]
  \centering
  \sidecaption
  \includegraphics[width=12cm]{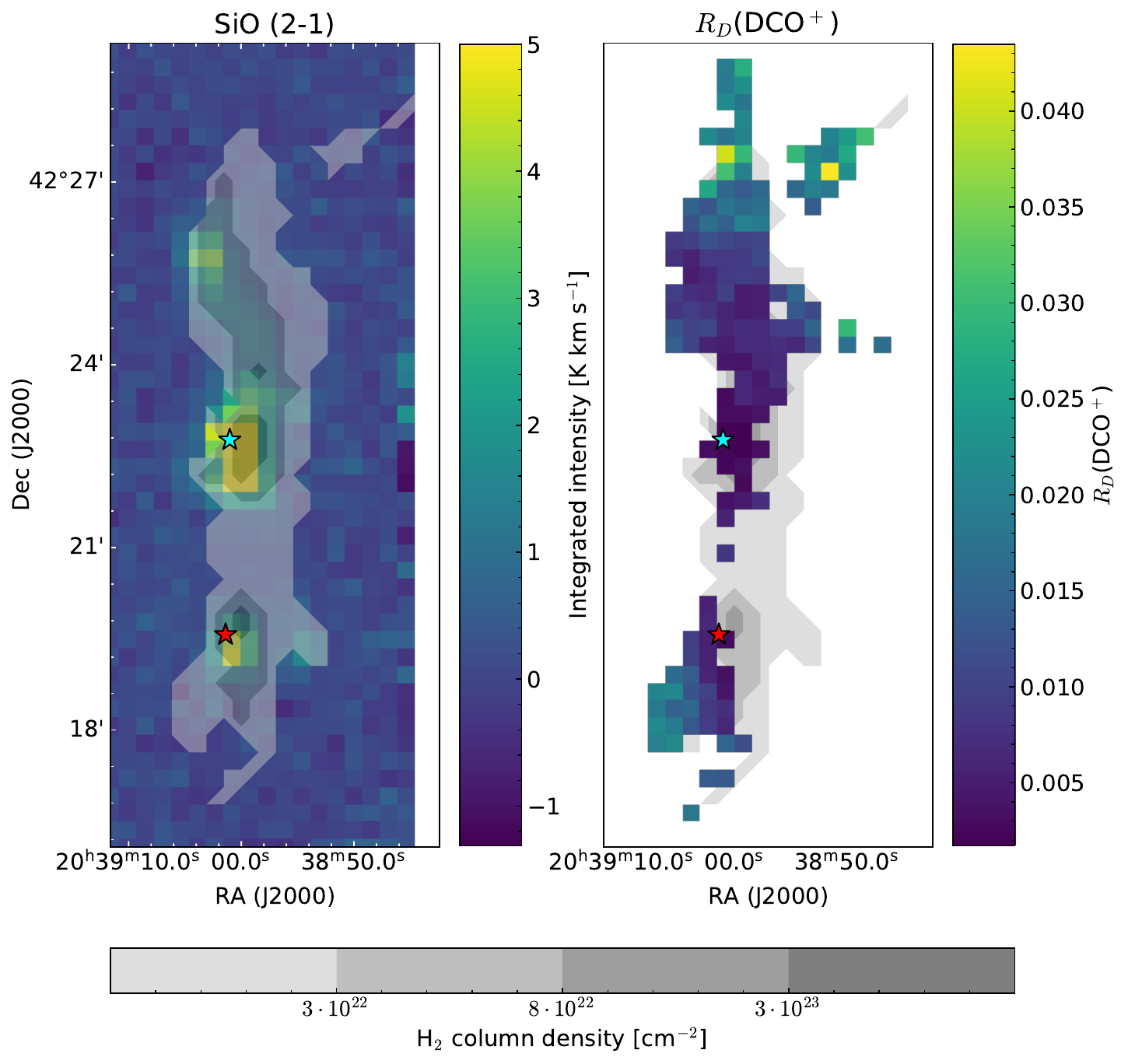}
  \caption{Distribution of SiO ($2-1$) integrated intensity (left panel) and  $R_D$(\ce{DCO+}) (right panel) the DR21 filament. The grey scale shows the N(H$_2$) column densities, where darker regions indicate higher column densities. The stars show DR21 Main (red) and DR21(OH) (blue).}
  \label{fig:SiOEmission}
\end{figure*}
We investigate the influence of different physical parameters and, in particular, $N$(\ce{H2}) on the $R_D$(\ce{DCO+}). 
We summarize here the underlying chemistry for the formation of \ce{DCO+}.

The formation of \ce{DCO+} and \ce{HCO+} are most efficient in the lower temperature regime ($T_{\rm kin}<20~$K). The primary formation pathway of \ce{DCO+} is the following gas-phase reaction \citep{Albertsson2013}:
\begin{equation}
  \ce{H2D+} + \ce{CO} \rightarrow \ce{DCO+} + \ce{H2}.
  \label{eq:DCOform}
\end{equation}
The formation pathway of the main isotopolog also requires CO:
\begin{equation}
  \ce{H3+} + \ce{CO} \rightarrow \ce{HCO+} + \ce{H2}.
  \label{eq:HCOform}
\end{equation}
In the low temperature ($T_{\mathrm{dust}}< 20$ K) and high density regime ($n > 10^{4}$~cm$^{-3}$, where CO is frozen onto the dust grains, the sparse amounts of CO available in gas-phase reacts with \ce{H2D+}. 
As CO is sublimated from the ices, CO reacts with \ce{H3+}, depleting the reservoir to form \ce{H2D+} from \ce{H3+}.
Consequently, $R_D$(\ce{DCO+}) peaks in $T_{\mathrm{gas}} \leq 20$ K and decreases with increasing temperature \citep{Aikawa2012}.
The formation of \ce{DCO+} and \ce{HCO+} decreases as \ce{H3+} is destroyed before forming \ce{H2D+} \citep{Albertsson2013}. 
Using a time-dependent chemical model, \cite{Albertsson2013} showed a thermal dependence of \ce{DCO+} in the lower temperature regime ($\sim 30$ K). 
Once the critical temperature for \ce{DCO+} formation is reached, the $R_D$(\ce{DCO+}) begins to decrease to the elemental abundance of D/H-ratio ($\sim 10^{-5}$) \citep{Cooke2018}.
A simple thermal dependence of $R_D$(\ce{DCO+}) in low ionization fraction regions, following the temperature dependence of its precursors, has been established \citep{Herbst1982, Anderson1999}. 

Elevated temperatures driven by shocks can cause the non-thermal sublimation or sputtering of CO from dust grains.
As a result, shocks can also suppress the formation of \ce{DCO+}, discussed in depth in Sect.~\ref{subsec:shocks}. 
Within the DR21 filament, an environment teeming with varied degrees of star formation activity along its structure \citep{Schneider10,Hennemann}, regions like DR21 Main and DR21(OH) stand out as discussed in Sect.~\ref{Sec:Results}. 
The \hii~region in DR21 Main, from which the prominent bipolar-shaped outflow originates, drives shock chemistry in its surroundings \citep[see, for example,][]{Skretas2023}. 
The latter region, DR21(OH), where the lowest $R_D$(\ce{DCO+}) value is observed, also exhibits clear signatures of shock chemistry \citep[see][]{Godard2012}. 
Therefore, the shocks present in the vicinity of these \hii~regions may play a vital role in the chemistry of $R_D(\ce{DCO+})$.

For a complete understanding of the chemistry of \ce{DCO+} in DR21, one must, in addition to the environment specific effects suppress its formation, also take into account the relevant destruction pathways. 
The primary pathway leading to the destruction of \ce{DCO+} and \ce{HCO+} is dissociative recombination:
\begin{equation}
    \ce{DCO+} + e^- \rightarrow \ce{CO} + \ce{D}
    \label{eq:DCOdestr}
\end{equation}
and
\begin{equation}
    \ce{HCO+} + e^- \rightarrow \ce{CO} + \ce{H}
    \label{eq:HCOdestr}
\end{equation}
As a result, a higher electron fraction accelerates the destruction of \ce{HCO+} more rapidly than that of \ce{DCO+}, thereby influencing the molecular isotopic ratio \citep{Albertsson2013}.

In the following sections, we discuss in more detail the influence of shocks and photodissociation on the chemistry of $R_D$(\ce{DCO+}) within the DR21 filament. 

\subsubsection{Effect of shocks}
\label{subsec:shocks}
In this section, we investigate the impact of shocks on the chemistry of $R_D(\ce{DCO+})$. 
As discussed above, the release of CO into the gas-phase readily destroys \ce{H3+} and, in turn, suppresses the formation of \ce{H2D+} and \ce{DCO+} (see Eq.~\eqref{eq:DCOform} and Eq.~\eqref{eq:HCOform}). 
The dominant pathway leading to the desorption of CO from dust grains at lower temperatures is via non-thermal desorption driven by shocks. 

Multiple sources of shocks are found in DR21, e.g. outflows, star-forming regions and cloud-cloud interactions. 
The DR21 filament harbors one of the most extreme outflow, by mass and size, in our Milky Way namely DR21 Main \citep{Skretas2023}.
Furthermore, the DR21 filament was proposed to be the result of cloud-cloud collisions \citep{Dickel1978, Dobashi2019}. Recent observations of ionized carbon \citep{Schneider2023, Bonne2023} have refined this scenario, revealing this region to be more complex with several molecular clouds and their \hi~envelopes interacting.

A commonly used molecular tracer of shocks is SiO, formed by the non-thermal sputtering of Si from dust which reacts with gas phase O \citep{Pintado1997}. 
The specific use of SiO as a shock tracer has been demonstrated through observations and models.
Observations show an increase in SiO abundances from quiescent, previously shocked, regions to dense, active regions as the center of our Milky Way \citep{Amo-Baladr2011, Rybarczyk2023}.
Theoretical models by \cite{Kelly2017} show the abundance of SiO to increase significantly (often by an order of magnitude) in regions where it encounters both slow and fast shocks. 
In various environments, these studies demonstrate that the presence or enhancement of the SiO molecule serves as an indicator of shock chemistry.

Given that CO is sputtered from the dust grain mantles alongside SiO, the detection of significant amounts of SiO in the gas-phase indirectly traces the source of \ce{H2D+} suppression.
While it is non-trivial to infer the properties of the shocks in DR21, e.g. shock velocity and shock age, through abundance measurements of SiO, we compare the spatial extent of SiO emission in DR21 to decipher the role of shocks in $R_D(\ce{DCO+})$ chemistry.

Figure~\ref{fig:SiOEmission} shows the spatial distribution of the integrated intensity map of SiO (2--1), obtained with CASCADE, and $R_D(\ce{DCO+})$, alongside $N$(\ce{H2}) between $3 \times 10^{22}$~cm$^{-2}$ and $1 \times 10^{24}$~cm$^{-2}$ overlaid in greyscale. 
The increased SiO (2--1) emission suggests that it coincides with the regions of decreased $R_D(\ce{DCO+})$.
The regions with the highest values of $N$(\ce{H2}) ($> 8\times 10^{22}$~{cm}$^{-2}$) correspond to that of the massive dense cores, DR21 Main and DR21(OH) (see Fig.~\ref{fig:SiOEmission}) which coincides with the regions where the SiO (2-1) emission is the strongest. 
More importantly, from this comparison it is clear that the regions where the SiO (2--1) emission dominates is also spatially coincident with the regions where we estimate the lowest values for $R_D$(\ce{DCO+}). 
This lends credence to our hypothesis that shocks in the vicinity of DR21 Main and DR21(OH) play an important role in regulating the chemistry and hence the deuteration present in \ce{DCO+}. 
In addition, the attenuation of cosmic-rays in dense region such as DR21(Main) and DR21(OH) might diminish the abundance of \ce{H3+}, thus decreasing the \ce{HCO+} and \ce{DCO+} abundances in these regions. 
Testing the importance of cosmic-rays in regulating the $R_D$(\ce{DCO+}), however, is beyond the scope of this paper.

\begin{figure}[htbp!]
  \centering
  \includegraphics[width=\linewidth]{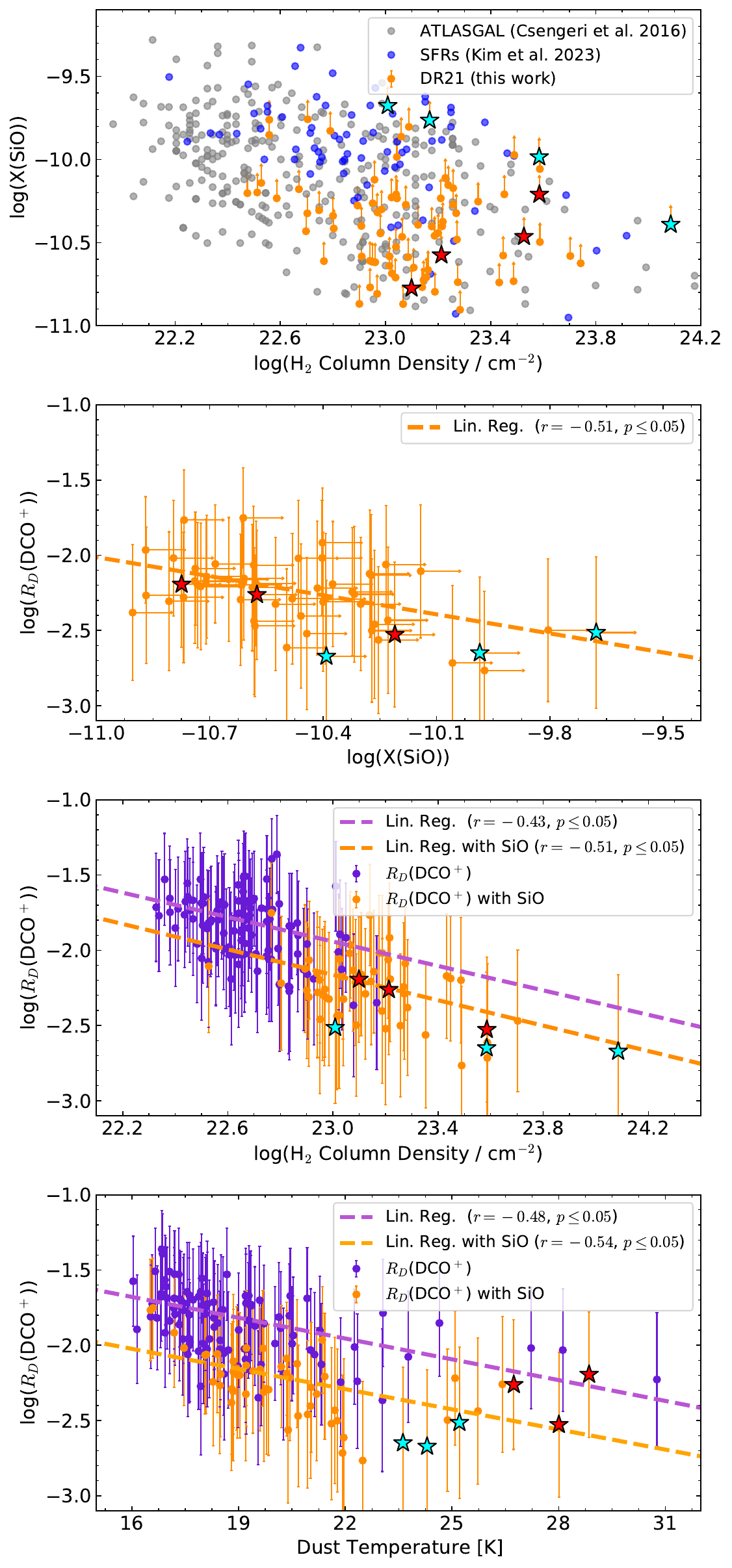}
  \caption{The abundance of SiO with \ce{H2} column density in the DR21 filament and ATLASGAL sources \citep{Csengeri2016} and SFRs \citep{Kim2023}, and $R_D$(\ce{DCO+}) with the abundance of SiO in the DR21 filament (first and second row, respectively). 
  The third row and fourth row shows the \ce{H2} column density and the dust temperature with $R_D$(\ce{DCO+}), respectively, where purple shows non-detection of SiO and orange shows the pixels where SiO was detected $\geq 4\sigma$.
  The red and blue stars mark the surroundings of DR21 Main and DR21(OH), respectively.}
  \label{fig:SiO_trends}
\end{figure}

In the top two panels of Fig.~\ref{fig:SiO_trends} we examine variations of SiO abundance with $N(\ce{H2})$. 
We find a similar trend as for APEX Telescope LArge Survey of Galaxy \citep[ATLASGAL;][]{Schuller2009} sources SiO measurements by \cite{Csengeri2016}, and the SiO $J=1-0$ survey toward massive SFRs \citep{Kim2023}.
These authors find that the embedded \hii~regions show a higher abundance compared with that of younger sources, where the shock conditions change with the massive clumps or with the age of the clump.
Similarly, we find abundances of SiO in the DR21 filament as in the shocked ATLASGAL sources and SFRs at similar \ce{H2} column densities.

We find that the regions toward which we detect SiO (2--1) emission at rms noise levels $\geq 4 \sigma$ correspond those regions with consistently low values of $R_D(\ce{DCO+})$ ($<0.03$). 
The second panel in Fig. \ref{fig:SiO_trends} shows a decreasing trend of $R_D(\ce{DCO+})$ with the abundance of SiO. However, this decrease occurs due to \ce{H2} column density, as the SiO abundance is independent of the \ce{H2} column density in the DR21 filament.
The Pearson correlation coefficient of $R_D$(\ce{DCO+}) with \ce{H2} column density and $T_{\rm dust}$, where SiO is detected, is $r = -0.51$ and $r = -0.54$, respectively. 
While a definitive conclusion regarding the state of shocked gas in DR21 cannot be reached, the spatial correlation between the shock tracer and the decreased $R_D(\ce{DCO+})$ strongly suggests that the shocks are influencing the chemistry of $R_D(\ce{DCO+})$.

\subsubsection{Effects of photodissociation by UV radiation}
\label{subsec:photodissociation}

In this section, we investigate the impact of photodissociation on the chemistry of $R_D(\ce{DCO+})$. 
The interface in the interstellar medium, where the chemistry is primarily influenced by UV photons, is known as a photodissociation region \citep[PDR;][]{Hollenbach1997}.
At the forefront of a PDR, \ce{H2} molecules undergo photodissociation. 
As the \ce{H2} column density increases (toward regions with higher extinction), the low column density ($N$(H) $\leq 2 \times 10^{20}$ cm$^{-2}$), ionized medium transitions to the molecular, high density ($N$(H) $\geq 2 \times 10^{22}$ cm$^{-2}$) medium.
Despite only having modest abundances in the ISM, HCO is efficiently formed in PDRs via FUV \citep{Schilke2001, Gerin2009}:
\begin{equation}
    \ce{H2CO} + \ce{photon} \rightarrow \ce{HCO} + \ce{H}
\end{equation}
and
\begin{equation}
    \ce{O} + \ce{CH2} \rightarrow \ce{HCO} + \ce{H}\,,
\end{equation}
where the abundance of this rare molecular species becomes comparable with that of \ce{H^13CO+}.
Furthermore, a high abundance of \ce{HCO} in the gas phase can be attained by releasing \ce{HCO} from grain mantles, before larger molecules form there, for example, \ce{H2CO}, \ce{CH3OH} and other complex organic molecules \citep{Gerin2009}. 
This early release of \ce{HCO} to the gas-phase can occur in regions with lower $N$(\ce{H2}) that are exposed FUV radiation.
Photodissociation is an obvious process associated with the well-studied \hii-region DR21 Main and DR21(OH). The outer parts of DR21 dominantly affected by the interstellar UV radiation \citep{Yamagishi2018}.

The column density ratio of HCO and \ce{H^13CO+} and the abundance of the latter, have been found to be increased in \hii~regions when compared to regions devoid of HII regions or showing minimal signs of star formation activity \citep{Schilke2001, Gerin2009, Kim2020}.
$N$(HCO)/$N$(\ce{H^13CO+}) $\geq 1$ and $X$(HCO)$\geq 10^{-10}$ reflects high degrees of photodissociation, and is indicative of ongoing FUV chemistry \citep{Gerin2009,Kim2020}. 

\begin{figure}[htbp!]
  \centering
  \includegraphics[width=\linewidth]{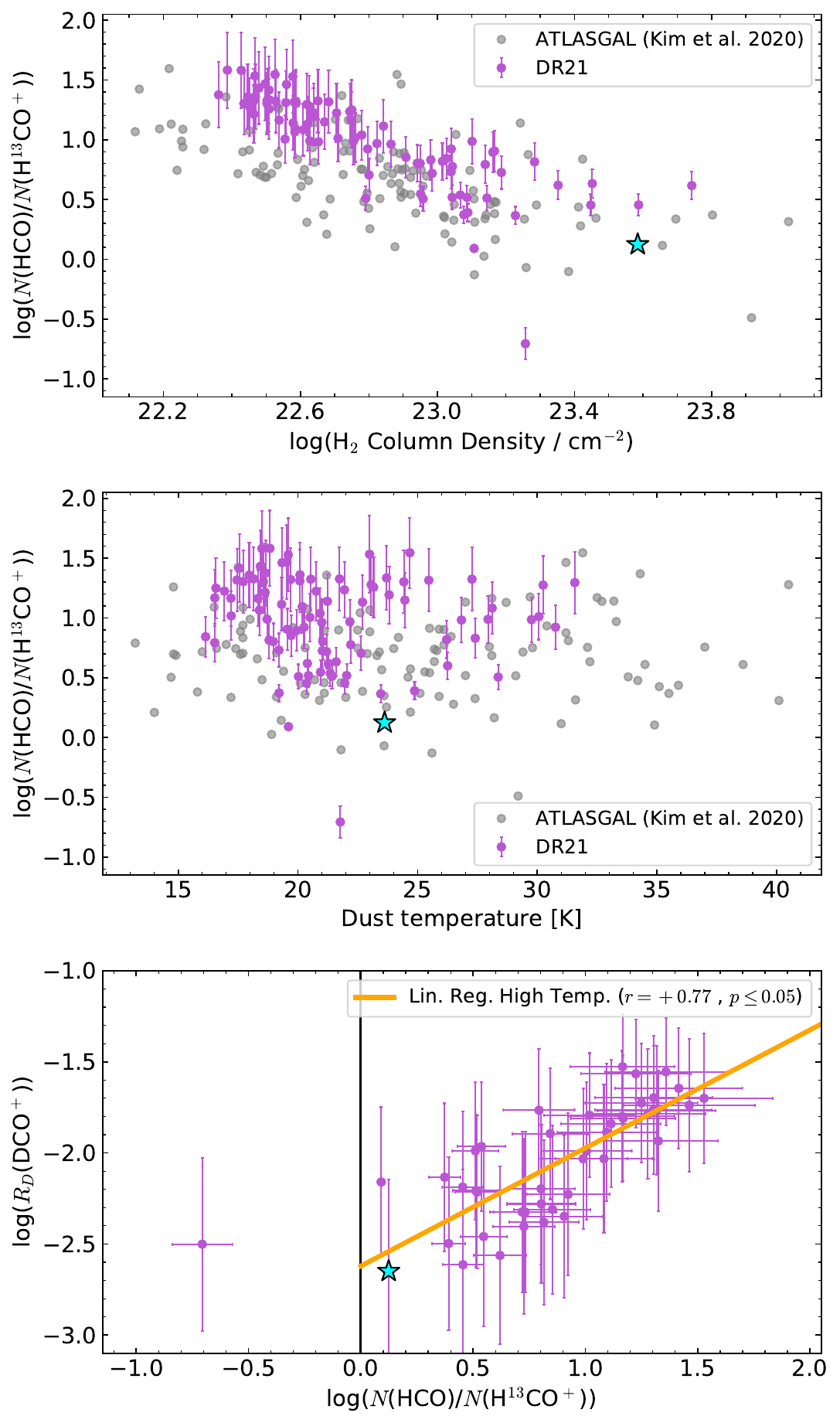}
  \caption{$N$(HCO)/$N$(\ce{H^13CO+}) vs. $N$(H$_2$) (top panel) and dust temperature (middle panel). The bottom panel shows the correlation of $N$(HCO)/$N$(\ce{H^13CO+}) with $R_D$(\ce{DCO+}). The purple points represent pixels in the DR21 filament above $4\sigma$ and the grey point shows the ATLASGAL sources from \cite{Kim2020}. The black vertical line represents the indicator of ongoing FUV chemistry $\geq 1$ \citep{Gerin2009, Kim2020}. 
  The blue star mark the surroundings of DR21(OH), where \ce{HCO}, \ce{H^13CO+} and \ce{DCO+} are detected.}
  \label{fig:ionizationHCO}
\end{figure}

In Fig.~\ref{fig:ionizationHCO}, we investigate the variation of FUV radiation fields by examining the column density ratio of $N(\ce{HCO})$ and $N(\ce{H^13CO+})$ in the DR21 filament.
The emission maps and column densities of \ce{HCO} are described in depth in App.~\ref{appendix:emissionmaps}.
In all panels of Fig.~\ref{fig:ionizationHCO}, the surroundings of DR21(OH) is marked with a blue star. 
Unfortunately, we did not detect \ce{HCO}, \ce{H^13CO+}, and \ce{DCO+} in any of the surroundings of DR21 Main, and we only detected one pixel with the three molecules simultaneously detected in the vicinity of DR21(OH).
The upper panel of Fig. \ref{fig:ionizationHCO} shows a decreasing trend of the $N$(HCO)/$N$(\ce{H^13CO+}) ratio with \ce{H2} column density in the DR21 filament compared with values of this ratio determined for molecular clumps found in ATLASGAL, taken from \citep{Kim2020}.   
The observed trend is mainly driven by the decreasing abundance of \ce{HCO} as a function of increasing \ce{H2} column density.  
The middle panel of Fig. \ref{fig:ionizationHCO} shows no variation of $N$(HCO)/$N$(\ce{H^13CO+}) with dust temperature neither in DR21 nor the ATLASGAL clumps. 
No correlation between $N$(HCO)/$N$(\ce{H^13CO+}) and $T_{\rm dust}$ implies that thermal desorption does not play a role in the formation of HCO \citep{Kim2020}.

The lower panel of Fig. \ref{fig:ionizationHCO} shows a correlation between $N$(HCO)/$N$(\ce{H^13CO+}) and $R_D$(\ce{DCO+}) with a Pearson correlation coefficient of $r = 0.77$. 
This increasing trend of $R_D$(\ce{DCO+}) with $N$(HCO)/$N$(\ce{H^13CO+}) starts at the point at which the abundances of \ce{DCO+} and \ce{HCO} start to decrease, that is where FUV driven chemistry commences \citep{Gerin2009, Kim2020}. 
These conditions can be found as the photodissociation degree decreases. 
As the \ce{H2} column density increases, \ce{H2} self-shielding becomes effective resulting in the decrease of the \ce{HCO+} destruction (see Eq.~\eqref{eq:HCOdestr}).
The decreasing trend of $R_D$(\ce{DCO+}) with $N$(\ce{H2}) is caused by the decrease of the \ce{HCO+} destruction. 

\begin{figure}[htbp!]
  \centering
  \includegraphics[width=\linewidth]{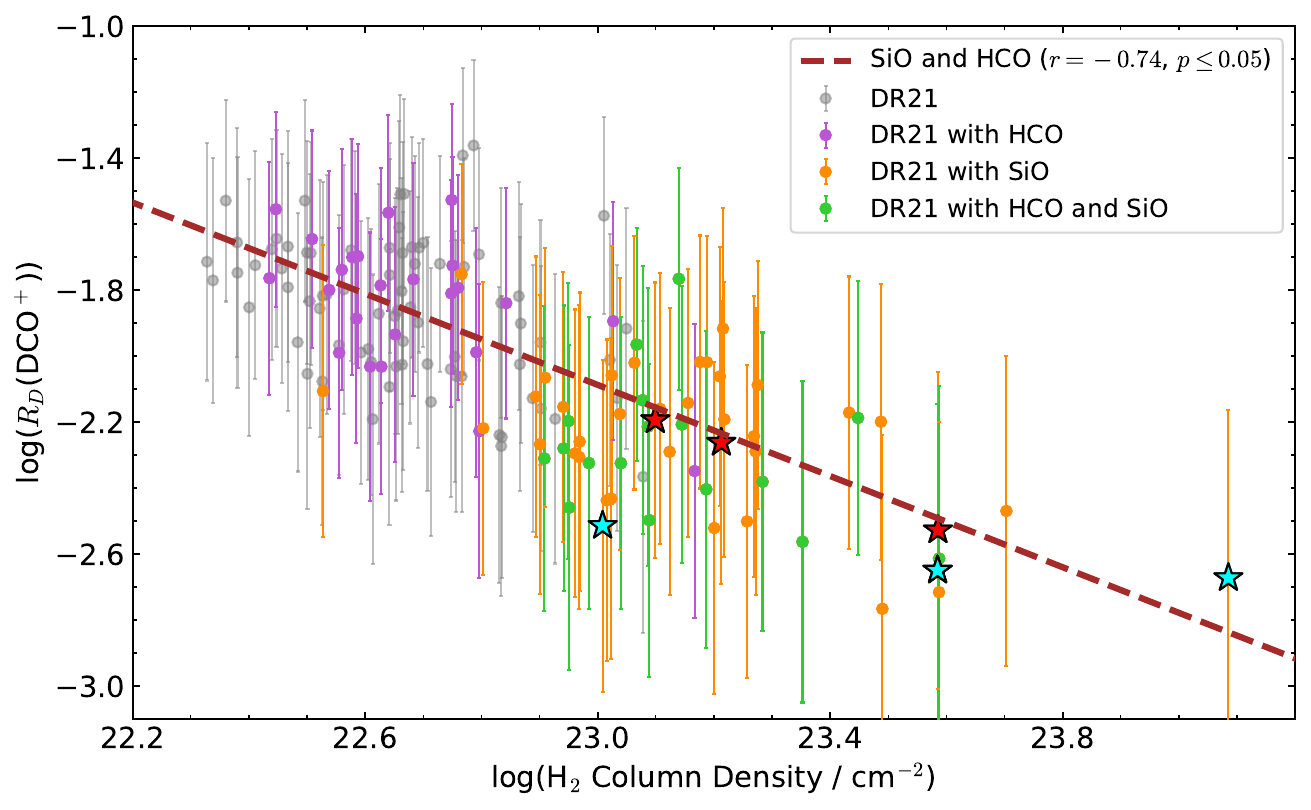}
  \caption{$R_D$(\ce{DCO+}) with column density, where grey are all measured pixels in DR21, purple are $R_D$(\ce{DCO+}) where $N$(HCO)/$N$(\ce{H^13CO+}) is measured, orange is where $X$(SiO) is measured and green is where $N$(HCO)/$N$(\ce{H^13CO+}) and $X$(SiO) intersect. The linear fit, where SiO and HCO are detected, is shown in dashed brown.
  The red and blue stars mark the surroundings of DR21 Main and DR21(OH), respectively.}
  \label{fig:DCOHCO_HCO_SiO}
\end{figure}

To summarize the discussion on the influence of shocks and ionization, the two physical processes affect $R_D$(\ce{DCO+}) in different ways. 
\ce{HCO+} is destroyed faster than \ce{DCO+} through dissociative recombination. 
The regions of $R_D$(\ce{DCO+}) with column density, where HCO and/or SiO are present, are marked with green in Fig. \ref{fig:DCOHCO_HCO_SiO}. 
We find a Pearson correlation coefficient $r = -0.79$ of $R_D$(\ce{DCO+}) with \ce{H2} column density where HCO is detected in Sect.~\ref{subsec:photodissociation}. 
These are found at the lower \ce{H2} column densities.
On the contrary, at higher \ce{H2} column densities, where SiO is detected, we find a Pearson correlation coefficient $r = -0.51$ of $R_D$(\ce{DCO+}) with \ce{H2} column density in Sect.~\ref{subsec:shocks}.
As CO reacts with \ce{H3+}, the production of \ce{H2D+} is halted, successively decreasing the formation of \ce{DCO+}. 
This suggests that FUV-chemistry affects the lower column densities and shock-chemistry affects the higher column densities in DR21, overall resulting in the strong decrease of $R_D$(\ce{DCO+}) ($r = 0.74$). 

\subsection{Impact of physical conditions on DNC and DCN}
\label{sec:Temperature}
In  Sect.~\ref{sec:AnalysisMaps}, it is observed that DNC and DCN are indicative of different physical conditions compared to \ce{DCO+}. 
Specifically, they are detected in warmer and denser regions within the DR21 filament. Figure~\ref{fig:phys_distribution} illustrates that these isomers trace similar gas distributions. In this section, we aim to investigate the influence of temperature on the ratio of DNC to DCN.

Figure~\ref{fig:Beautiful_corr} shows that $R_D(\ce{DNC})$ and $R_D(\ce{DCN})$ have distinct dependencies on $T_{\rm dust}$. 
The formation pathways of these isomers have significant differences in their thermal dependence. 
In the lower temperature regime ($T_{\mathrm{dust}} \leq 30$ K), DNC is predominantly formed at a higher rate through deuterium reacting with the primary isotopologs, HCN and HNC. Surprisingly, in this temperature range, DCN formation occurs through the deuterium transfer of HNC \citep{Turner2001}. Conversely, DNC and DCN efficiently form with deuterated light hydrocarbons, such as \ce{CH2D+}, at higher temperatures \citep[$\sim 80$ K;][]{Albertsson2013}.
At lower gas temperatures ($T_{\mathrm{gas}}\leq 30$ K), DCN exhibits lower formation efficiency, as it is formed through \ce{CH2D+}, compared to DNC, formed through \ce{H2D+} \citep{Turner2001, Roueff2007}. 
As for DNC, its formation efficiency decreases with increasing temperature since the O-atom becomes more effective in destroying this isomer \citep{Gerner2015}. However, the current study does not explore these higher temperatures as the maximum dust temperature and kinetic gas temperature measured were $\leq 31$ K.

\begin{figure}[htpb!]
  \centering
  \includegraphics[width=\linewidth]{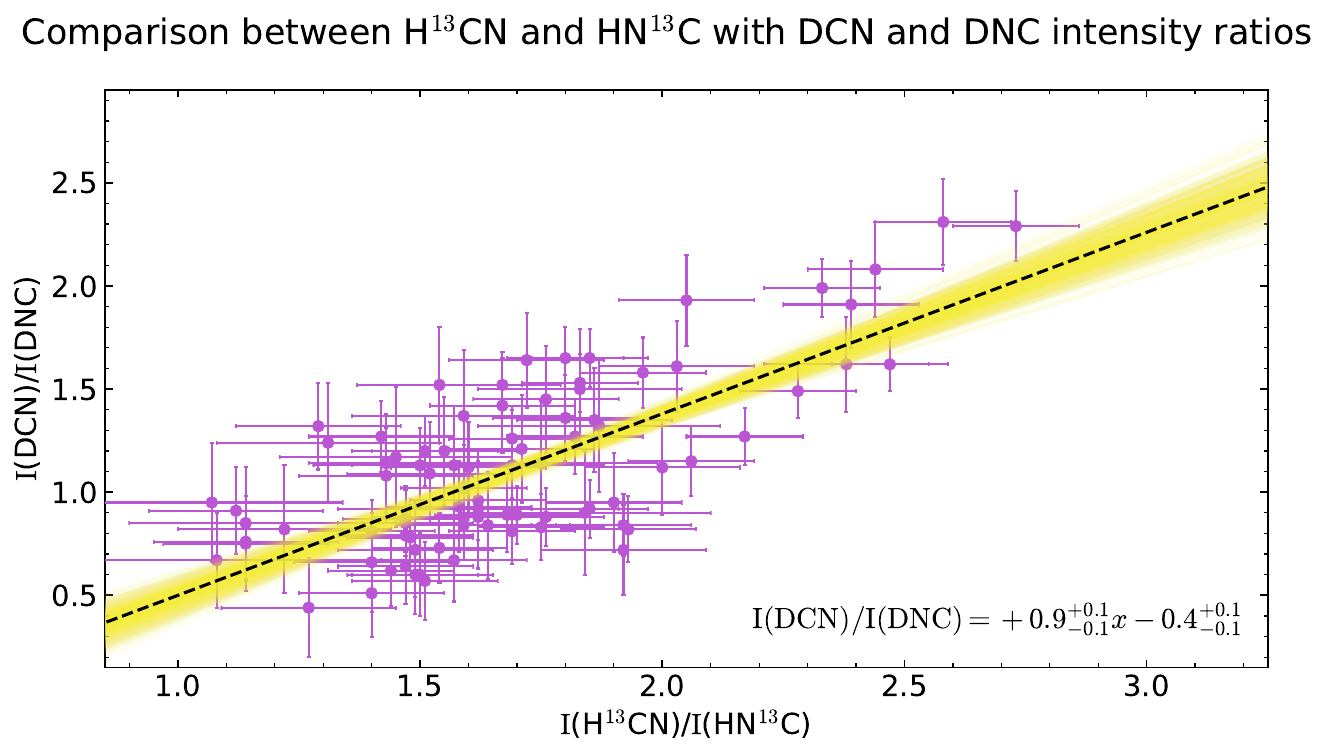}
  \caption{The integrated intensity ratio of DCN and DNC with the ratio of H$^{13}$CN and HN$^{13}$C of optically thin region (orange), as a proxy for the kinetic gas temperature following \cite{Hacar2020} with a Pearson correlation of $r = 0.72$.}
  \label{fig:DCNDNC}
\end{figure}

HCN and HNC, the primary isotopologs of these isomers, have demonstrated the ability to trace the kinetic gas temperature (see Sect.~\ref{sec:coldensdet}).
The ratio of $I$(HCN)/$I$(HNC) at gas temperatures $T \leq 200$ K is controlled by the selective destruction of HNC \citep{Hacar2020}:
\begin{equation}
  \text{HNC} + \text{H} \rightarrow \text{HCN} + \text{H}
\end{equation}
and, 
\begin{equation}
  \text{HNC} + \text{O} \rightarrow \text{NH} + \text{CO}
  \label{eq:HNCO}
\end{equation}
The inter-conversion between HCN and HNC becomes efficient at $T >40$~K. 
At low temperatures,  HCN and HNC are formed through ion-molecule chemistry with almost equal probability \citep{Herbst2000,Albertsson2013}.
A similar effect may occur for DCN and DNC, where an increase in $T_{\mathrm{kin}}$ is necessary to initiate the conversion of DNC to DCN through reactions with free atomic D in the gas.

Figure~\ref{fig:DCNDNC} shows the intensity ratio comparisons of $I$(DCN)/$I$(DNC) with $I$(\ce{H^13CN})/$I$(\ce{HN^13C}). 
In the optically thin regime of these ratios, $I$(\ce{H^13CN})/$I$(\ce{HN^13C}) is a proxy for the kinetic gas temperature \citep{Beuther2022, Pazukhin2022}. 
The optical depth of each molecule is presented in Fig.~\ref{fig:Tau_maps}, which is consistent with the optically thin assumption. 
The intensities are used to ensure independent ratio comparison, as the column densities are inherently dependent on the temperature calculated using $I$(\ce{H^13CN})/$I$(\ce{HN^13C}).
The D-isotopologs $I$(\ce{H^13CN})/$I$(\ce{HN^13C}) with a slope $a = +0.9_{-0.1}^{+0.1}$ and intercept of $b = -0.4_{-0.1}^{+0.1}$.
The temperature indicator represented by $I$(\ce{H^13CN})/$I$(\ce{HN^13C}) exhibits a moderate correlation with $I$(DCN)/$I$(DNC), implying that $I$(DCN)/$I$(DNC) is also a potential tracer of kinetic gas temperature.

\begin{figure}[htpb!]
  \centering
  \includegraphics[width=\linewidth]{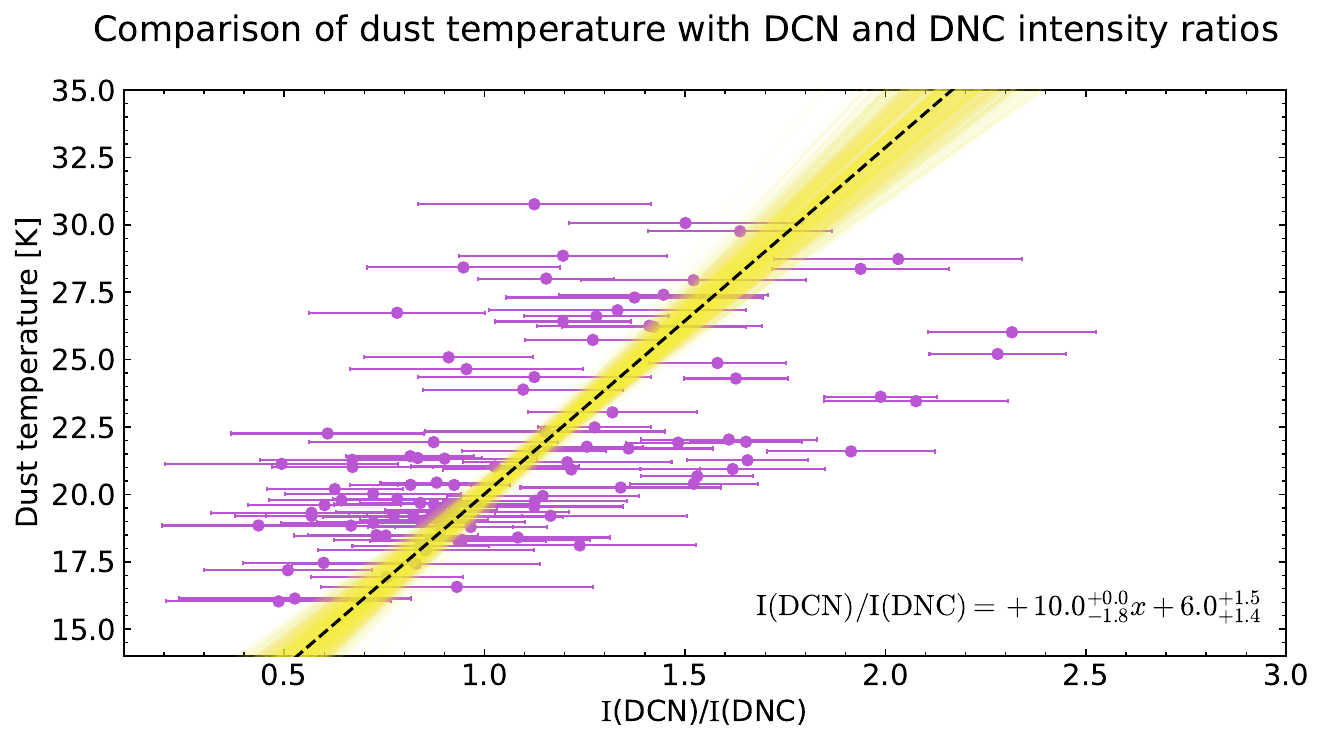}
  \caption{The dust temperature as a function of integrated intensity ratio of DCN and DNC with a Pearson correlation of $r = 0.58$.}
  \label{fig:DustDCNDNC}
\end{figure}

Figure~\ref{fig:DustDCNDNC} shows a plot of dust temperature as a function of $I$(DCN)/$I$(DNC), which displays a moderate correlation between
the dust temperature and the velocity-integrated intensity ratio of DCN and DNC.
We find a slope $a = +10$ and intercept of $b = +6$ between the dust temperature and $I$(DCN)/$I$(DNC). 
This correlation implies a kinetic gas temperature of $T_{\mathrm{dust}} = 10 \times (\mathrm{DCN/DNC}) + 6$. 
\cite{Hacar2020} reported a one-to-one correlation between dust temperature and the kinetic temperature derived from the intensity ratio of HCN to HNC. 
By applying the same approach to obtain kinetic gas temperature (Eq.~\eqref{eq:Tkin_calc1}), but using the intensity ratio of DCN to DNC instead and multiplying it by 10, we establish a one-to-one relationship between the kinetic gas temperature ($T_{\mathrm{kin}}$) derived from DCN/DNC and the dust temperature in DR21. 

\section{Conclusions}
\label{Sec:Conclusions}\label{sec:Conclusions}
In this paper, we investigated the distribution of deuterated molecules in DR21 as part of the \textbf{C}ygnus-X \textbf{A}llscale \textbf{S}urvey of \textbf{C}hemistry \textbf{A}nd \textbf{D}ynamical \textbf{E}nvironment. 
Using the IRAM 30-m telescope, observations of the ground-state of \ce{DCO+}, DNC and DCN at 3--4 mm and follow-up observations of higher $J$-transitions at 2 and 1.3 mm wavelength were carried out. 
The main results are summarized below:
\begin{itemize}
    \item[1.] The emission maps of \ce{DCO+}, DNC and DCN show clear morphological differences along the DR21 filament. \ce{DCO+} shows the most extended emission along the DR21 ridge, with several peaks along its structure. 
    DNC is also observed to have several peaks but less extended emission than \ce{DCO+}. DCN shows relatively compact emission toward the DR21 Main and DR21(OH) star-forming clumps. The morphological differences of these three molecules suggest local variations of physical parameters. 
    
    \item[2.] The column densities derived from the optically thin lines from \ce{DCO+}, DNC and DCN, as well  H$^{13}$CO$^+$, HN$^{13}$C and H$^{13}$CN, together with the Herschel-based \ce{H2} column density, indicate  similar mean abundances of the deuterated molecules, on the order of $X(\text{D})\sim 6-9 \times 10^{-11}$. All three molecules are observed toward with higher \ce{H2} column densities.
    
    \item[3.] The deuteration degrees, $R_D$, of all three species hav similar values, between 0.010 and 0.014. $R_D$(\ce{DCO+}) and $R_D$(DNC) show a strong anti-correlation with dust temperature ($r < -0.5$) and all three show strong anti-correlation with kinetic gas temperature, the highest value of which ($r = -0.70$) is found for $R_D(\ce{DCO+})$
    
    The three ratios between the deuterated molecules, $N$(\ce{DCO+})/$N$(DNC) and $N$(\ce{DCO+})/$N$(DCN) show anti-correlations with the dust temperatures in the regime in which CO is not yet sublimated ($T_{\mathrm{dust}} \leq 22$ K). On the contrary, $N$(DCN)/$N$(DNC) shows an increasing correlation at $T_{\mathrm{dust}} \leq 22$ K. After the onset of thermal CO-sublimation, all three ratios are constant with dust temperature. 

    \item[4.] $R_D$(\ce{DCO+}) is decreasing with \ce{H2} column density. 
    We attempted to explain this decrease with ionization and shocks. At lower column densities, FUV chemistry, as traced by $N$(\ce{HCO})/$N$(\ce{H^{13}CO+}), is likely causing the rapid decrease by effectively destroying \ce{HCO+} faster than \ce{DCO+}. 
    $R_D$(\ce{DCO+}) decreases with an increasing column density of SiO, well known shock tracer.
    Ionization, shocks and the initial ortho-to-para H$_2$ ratio, although the latter is not quantified, appear to have an effect on the decrease of $R_D$(\ce{DCO+}). The relative influence of the degree of ionization or shocks affecting the decreasing $R_D$(\ce{DCO+}) is hard to discern solely based on the molecular content, as DR21 is a dynamic and complex filament with a powerful \hii-region, a substantial outflow, several star-forming regions in different evolutionary stages and cloud-cloud interactions. 

    \item[5.] The DCN/DNC-ratio was investigated as a possible kinetic gas temperature tracer. 
    At higher $N(\ce{H2})$, with the \ce{HN^13C} and \ce{H^13CN} lines being optically thin, a strong correlation is found. The obtained correlation is $I(\ce{DCN})/I(\ce{DNC}) = 0.9 \times I(\ce{H^13CN})/I(\ce{HN^13C}) - 0.4$. 
\end{itemize}

\begin{acknowledgements}
I.B.C. is a member of the International Max-Planck Research School at the universities of Bonn and Cologne (IMPRS).
This work is based on observations carried out under project number 145-19, 031-20 and 053-21 with the IRAM 30m telescope. IRAM is supported by INSU/CNRS (France), MPG (Germany) and IGN (Spain).
D.S. acknowledges support from the European Research Council under the
Horizon 2020 Framework Program via the ERC Advanced Grant Origins 83 24 28.
N.S. acknowledges support from the FEEDBACK-plus project that is
supported by the BMWI via DLR, Projekt Number 50OR2217.
N.C. acknowledges funding from the European Research Council (ERC) via the ERC Synergy Grant ECOGAL (grant 855130), from the French Agence Nationale de la Recherche (ANR) through the project COSMHIC (ANR-20- CE31-0009).
\end{acknowledgements}

\bibliographystyle{aa}
\bibliography{aa.bib}

\begin{appendix}

\section{CASCADE emission maps}
\label{appendix:emissionmaps}
\begin{figure*}[htbp!]
    \centering
    \includegraphics[width=\linewidth]{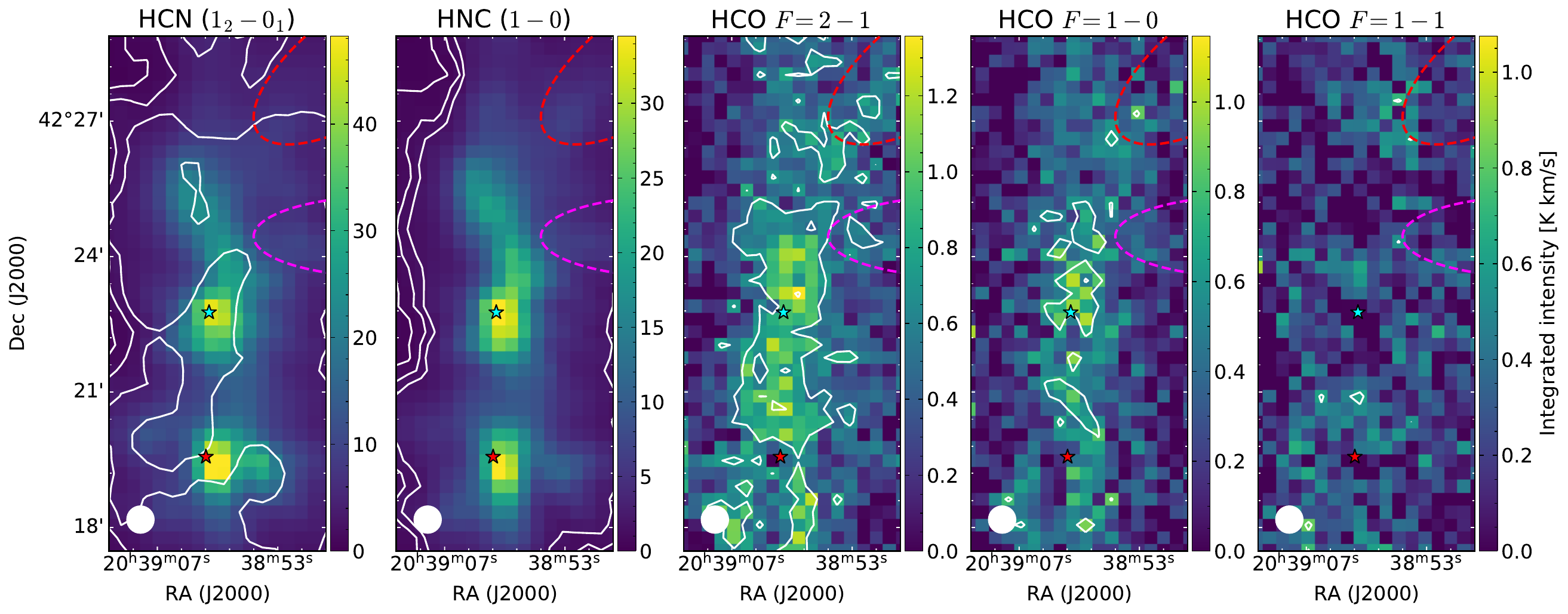}
    \caption{The ground-state emission of HCN, HNC and HCO observed with the CASCADE-program. 
    The white contours show the $4\sigma$,  $8\sigma$ and $12\sigma$ levels. The stars show DR21 Main (red) and DR21(OH) (blue), with the F1 and F3 sub-filaments marked in a red and magenta dashed curves.}
    \label{fig:CASCADEMainandiso}
\end{figure*}

Velocity-integrated intensity maps of the strong and extended emission of the main isotopologs HCN and  and HNC can be seen in two leftmost panels of Fig. \ref{fig:CASCADEMainandiso}. The emission in both species peak toward DR21 Main and DR21(OH).
The intensity ratio between these two molecules is used for kinetic temperature determinations, as described in Sect.~\ref{sec:coldensdet}.

The three right panels show the integrated intensity distributions of the three \ce{HCO} HFS components of this molecule's $N_{K_a,K_c} = 1_{0,1} - 0_{0,0}$ rotational line. These are the $J= 3/2 - 1/2, F= 2-1$, $J= 3/2 - 1/2, F= 1-0$ and $J= 1/2 - 1/2, F= 1-1$ lines. 
Spectroscopic information for the HCO HFS lines is presented in Table \ref{tab:HCO}.

Spectroscopic information for the the SiO $(2-1)$ lines is presented in Table~\ref{tab:HCO}. 
SiO (2--1) is covered as part of the CASCADE program. 
The integrated intensity map of SiO (2--1) is shown in Fig.~\ref{fig:SiOEmission}.

\begin{table*}[htbp!]
  \centering
    \caption{Spectroscopic information for the HCO and SiO lines covered by CASCADE.}
  \begin{tabular}{llccccc}
  \hline  \hline \\[-0.1mm]
  Species &  Transition & Frequency  & $E_{\rm up}/k$ & $A_{\rm ij}$ & $g_{\rm up}$ & Database  \\ 
   &  & [MHz] &[K] & [s$^{-1}$] & &    \\ \hline
  HCO &  $N_{K_a,K_c} = 1_{0,1} - 0_{0,0}$, $J= 3/2 - 1/2, F= 2-1$ & 86670.76 & 4.18 & 4.67(-06) & 5 & JPL  \\ [0.2cm]
  HCO &  $N_{K_a,K_c} = 1_{0,1} - 0_{0,0}$, $J= 3/2 - 1/2, F= 1-0$ & 86708.36 & 4.16 & 4.60(-06) & 3 & JPL \\ [0.2cm]
  HCO &  $N_{K_a,K_c} = 1_{0,1} - 0_{0,0}$, $J= 1/2 - 1/2, F= 1-1$ & 86777.46 & 4.18 & 4.60(-06) & 3 & JPL \\ [0.2cm] \hline \\ [-0.2mm]
  SiO, v=0-10 & 2 -- 1 & 86846.985 & 6.25 & 2.93(-05) & 5 & CDMS \\ [0.2cm] \hline
  \end{tabular}
  \tablefoot{The frequency, upper energy ($E_{\rm up}/k$), Einstein coefficient ($A_{\rm ij}$) and upper state degeneracy ($g_{\rm up}$) are marked. 
  The values for $A_{\rm ij}$ are written in the form of $x(-a) = x \times 10^{-a}$. The molecular are properties are from the JPL \citep{JplCat} and CDMS \citep{CDMS} spectroscopic catalogues, as noted.}
  \label{tab:HCO}
\end{table*}

\section{Optical depth}
\label{Appendix:OpticalDepth}
\begin{figure*}[htbp!]
    \centering
    \includegraphics[width=\linewidth]{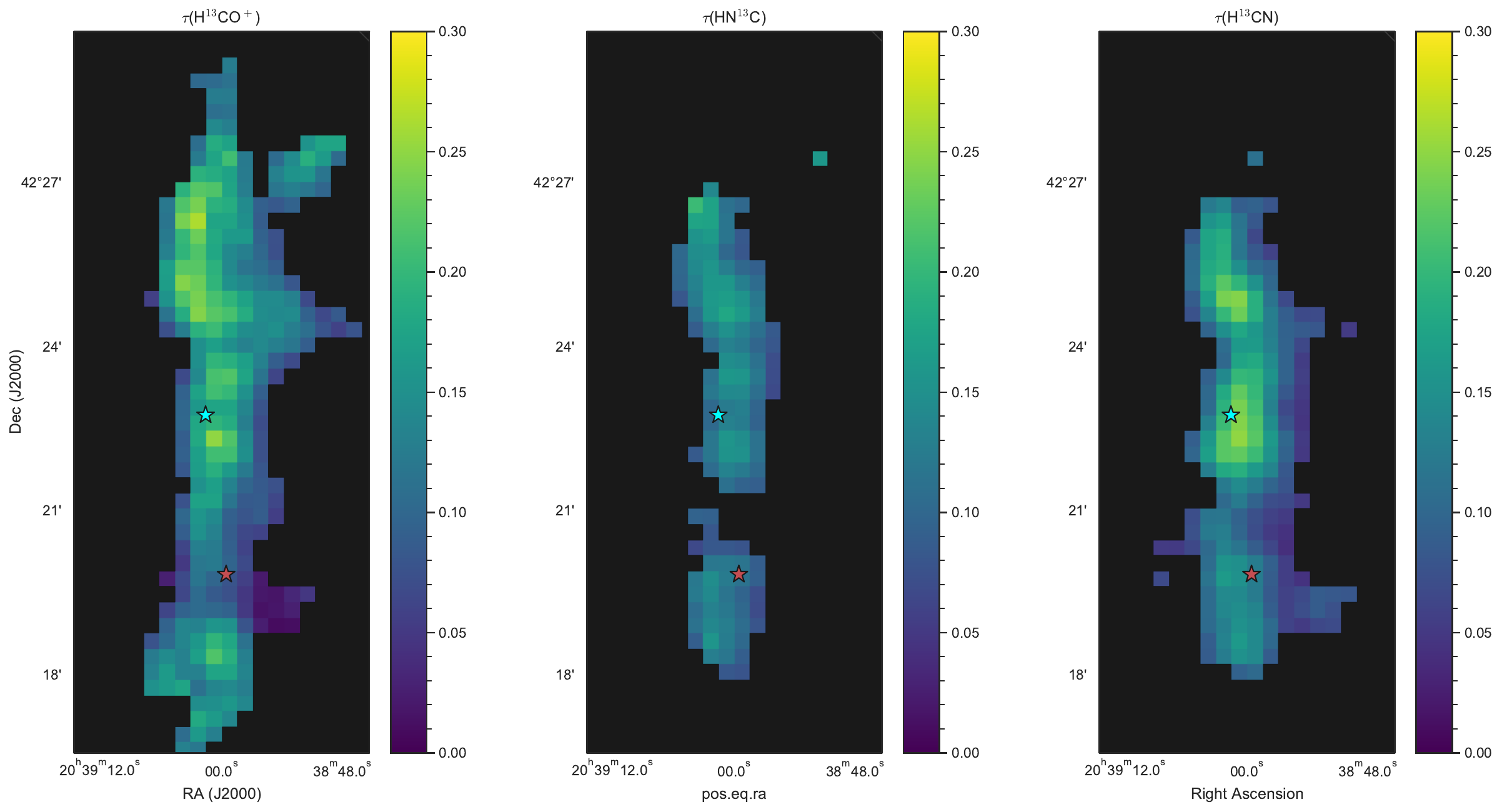}
    \caption{Left to right: The opacity maps of \ce{H^13CO+}, \ce{HN^13C} and \ce{H^13CN}. The stars mark the positions of DR21 Main (red) and DR21(OH) (blue).}
    \label{fig:Tau_maps}
\end{figure*}

The \ce{^12C} bearing main isotopolog lines of the molecules studied by us are optically thick and even the opacities of the $^{13}$C bearing isotopologs are expected to be non-negligible. The determination of the optical depth of the lines from each molecule relies on the ratio  of its \ce{^12C} bearing to its \ce{^13C} bearing isotopolog. Under the assumption that the beam filling factor, $f$, for each pixel is equal for both lines,
the ratio of the main beam brightness temperature of a line from the main   isotopolog 
$T_\mathrm{MB}^{\ce{^12C}}$ to that of the less common $\ce{^13C}$-bearing isotopolog, $T_\mathrm{MB}^{\ce{^13C}}$, is given by: 
\begin{equation}
\frac{T_\mathrm{MB}^{\ce{^12C}}}{T_\mathrm{MB}^{\ce{^13C}}}  \left( \upsilon \right)  = \frac{(T_\mathrm{ex}^{\ce{^12C}}-T_\mathrm{bg})(1 - \mathrm{exp}(-\tau_{\upsilon}^{\ce{^12C}})}
     {(T_\mathrm{ex}^{\ce{^13C}}-T_\mathrm{bg})(1 - \mathrm{exp}(-\tau_{\upsilon}^{\ce{^13C}})},
\end{equation}
where, $T_\mathrm{bg}$ is the temperature of cosmic microwave background (2.73~K). Further assuming that the frequencies and the excitation temperatures, $T_\mathrm{ex}$, of both lines are the same, results in an isotopic ratio of 68 for Cygnus-X (see Sect. \ref{sec:coldensdet}) and that the $^{\ce{^13C}}$ line is optically thin (so that $(1 - \mathrm{exp}(-\tau_{\upsilon}^{\ce{^13C}}) \approx \tau_{\upsilon}^{\ce{^13C}}$), this simplifies to:

\begin{equation}
0 \approx \frac{1 - \mathrm{exp}(-68\tau_{\upsilon}^{\ce{^13C}})}{\tau_{\upsilon}^{\ce{^13C}}} - \frac{T_\mathrm{MB}^{\ce{^12C}}}{T_\mathrm{MB}^{\ce{^13C}}} \left( \upsilon \right).
\end{equation}
From the roots of this equation, we can determine $\tau_{\upsilon}^{\ce{^13C}}$ using the Python package \texttt{scipy.optimize.newton\_krylov}. 
Fig. \ref{fig:Tau_maps} presents the pixel-by-pixel opacity map of the DR21 filament.

\section{Column density maps}
\label{Appendix:ColDens}
\begin{figure*}[htbp!]
    \centering
    \includegraphics[width=\linewidth]{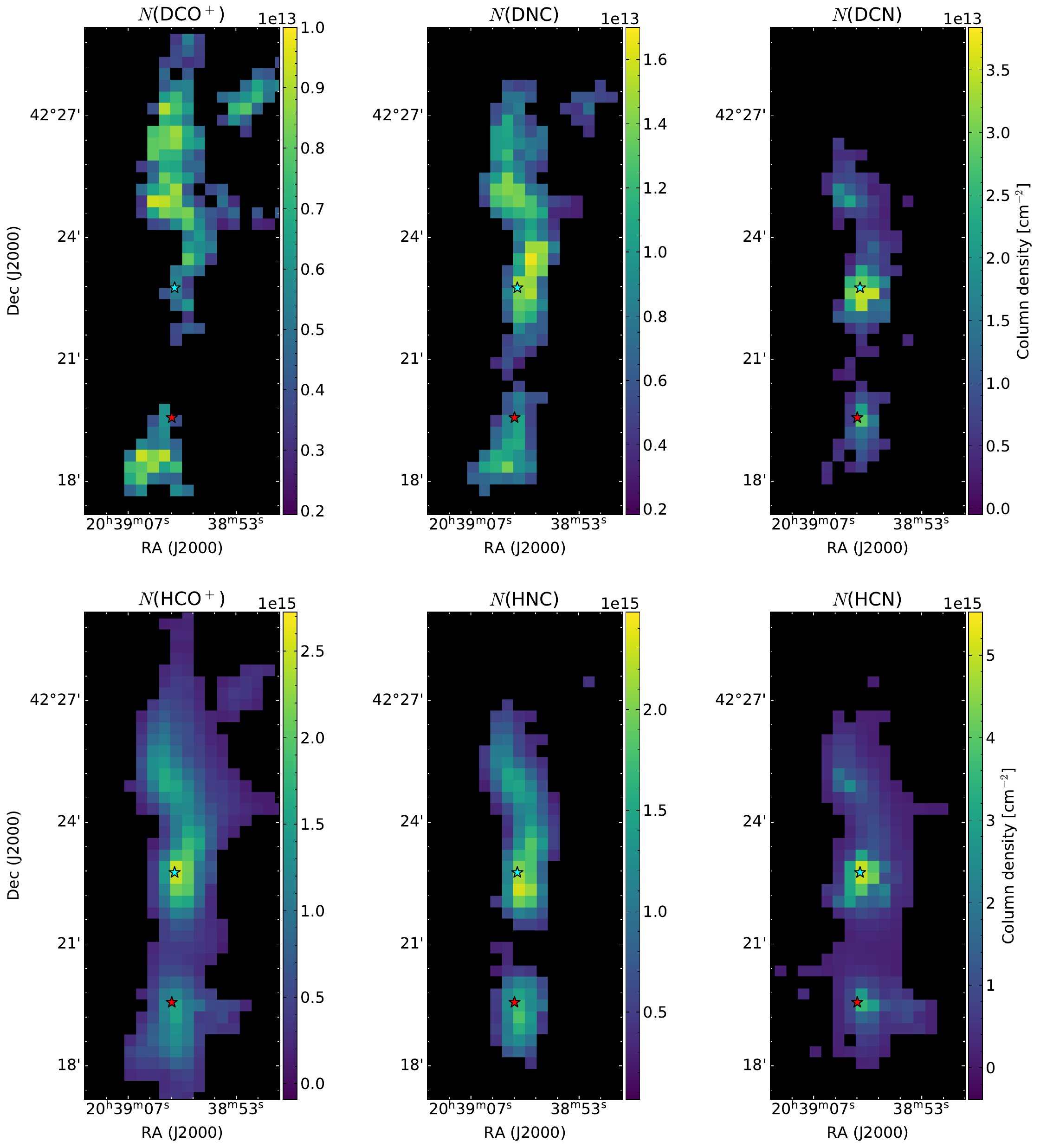}
    \caption{Top panel, left to right: The column density distribution maps of \ce{DCO+}, DNC and DCN and their main isotopologs \ce{HCO+}, HNC and HCN (bottom panel). The stars mark the positions of DR21 Main (red) and DR21(OH) (blue).}
    \label{fig:ColDensMap}
\end{figure*}

\begin{figure*}[htbp!]
    \centering
    \includegraphics[width=\linewidth]{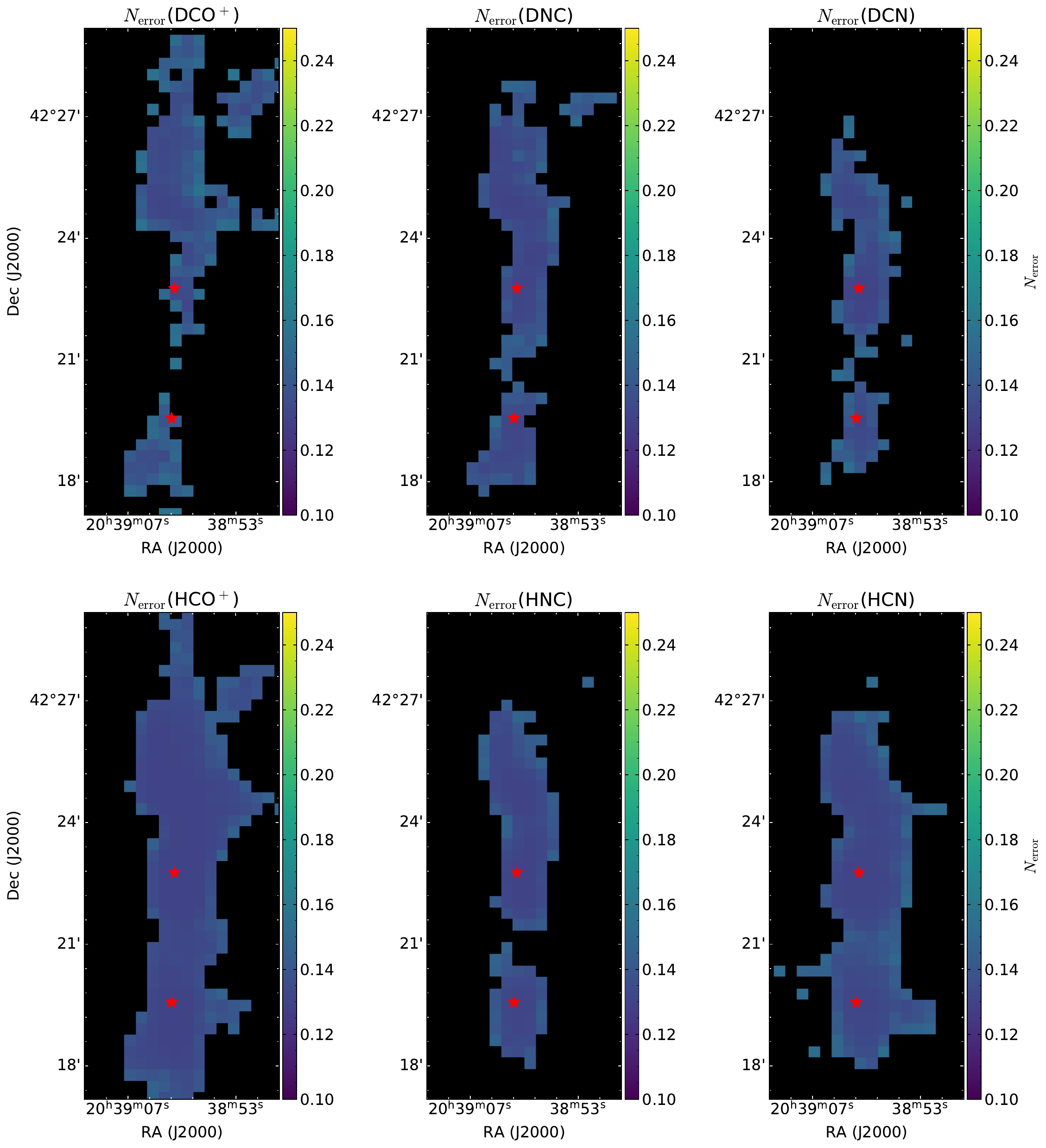}
    \caption{Top panel, left to right: The relative column density error distributions for \ce{DCO+}, DNC and DCN  and their main isotopologs \ce{HCO+}, HNC and HCN (bottom panel). The stars mark the positions of DR21 Main (red) and DR21(OH) (blue).}
    \label{fig:ColDensMap_error}
\end{figure*}

The distribution of the column densities derived in Sec. \ref{sec:coldensdet} are shown in Fig. \ref{fig:ColDensMap}.  
The column densities are determined using the \texttt{scipy.optimize.curve\_fit}, where each fitting of Eq.~\ref{eq:RotDiag} results in the best fit of $N_{\rm u}/g_{\rm u}$ and the respective fit-error. The column density maps for \ce{DCO+}, DNC and DCN, and the respective main isotopologs \ce{HCO+}, HNC and HCN are presented in Fig.~\ref{fig:ColDensMap} with the relative error maps presented in Fig. \ref{fig:ColDensMap_error}. 
The column density for the main isotopologs are calculated using the optically thin \ce{^13C}-isotopologs and assuming \ce{^12C}/\ce{^13C} $= 68$.

\section{N(XD)/N(YD)}
\begin{figure*}[htbp!]
  \centering
  \includegraphics[width=\linewidth]{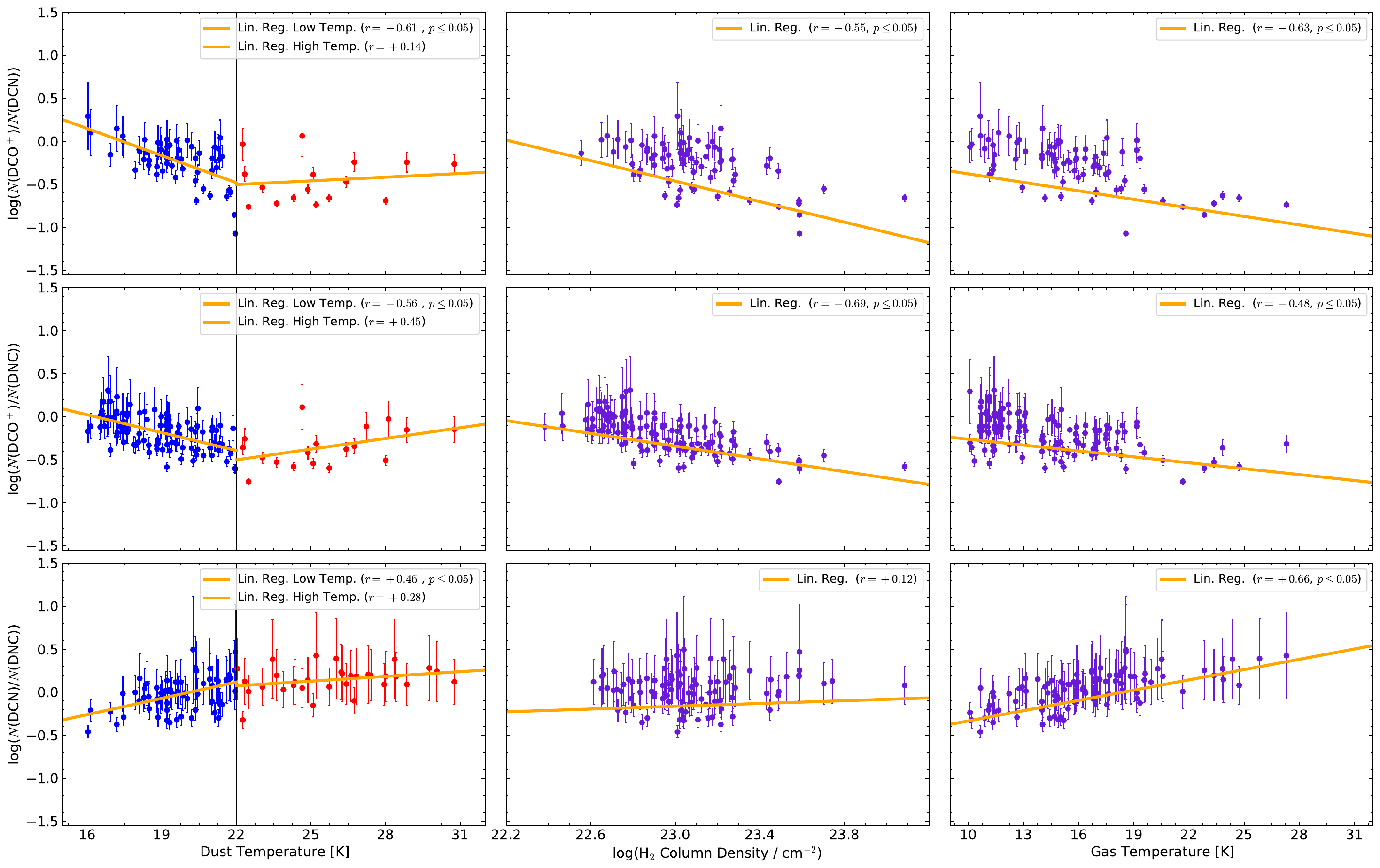}
  \caption{Deuterated molecular ratios with respect to the dust temperature (left) and column density (center) and kinetic gas temperature (right) of $N$(\ce{DCO+})/$N$(DCN) (top), $N$(\ce{DCO+})/$N$(DNC) (middle) and $N$(DCN)/$N$(DNC) (bottom). An onset of CO sublimation from grains is expected from $T_{\mathrm{dust}} \sim 22$ K, with blue data points showing the CO freeze-out regime.}
  \label{fig:DDwithTempandColD}
\end{figure*}

The dust temperature, \ce{H2} column density and kinetic gas temperature dependence of the $N$(XD)/$N$(YD) ratios are presented in Fig. \ref{fig:DDwithTempandColD}. 
The comparison of column density ratios with $T_{\rm dust}$ are presented in the left plots. 
Blue data points with $T_{\rm dust}$ below 22~K exhibit considerable temperature dependencies; $r=-0.61$ for log($N$(\ce{DCO+})/$N$(DCN)) and $r = -0.56$ for log($N$(\ce{DCO+})/$N$(DNC)).
A negligible dependence of log($N$(DCN)/$N$(DNC)) with temperature was found ($r = +0.46$).
On the contrary, all three ratios show no correlation with dust temperature in the higher temperature regime ($22~\ce{K} \leq T_{\mathrm{dust}} \leq 31$ K). 
The Pearson correlation coefficients of the higher temperatures are $r=+0.14$ for log($N$(\ce{DCO+})/$N$(DCN)), $r = +0.45$ for log($N$(\ce{DCO+})/$N$(DNC)), and $r = +0.28$ for log($N$(DCN)/$N$(DNC))) with $p > 0.05$. 

The $N$(XD)/$N$(YD) trends with \ce{H2} column density are $r = -0.55$ for log($N$(\ce{DCO+})/$N$(DCN)), $r = -0.69$ for log($N$(\ce{DCO+})/$N$(DNC)) and $r = +0.12$ for log($N$(\ce{DCN})/$N$(DNC)). 
This strong dependence is also reflected in the $R_D$(\ce{DCO+}) and $R_D$(DNC) ratio seen in the top left and middle of Fig. \ref{fig:Beautiful_corr}, respectively. The strong anti-correlation of $R_D$(\ce{DCO+}) is discussed further in Sect. \ref{sec:DCOHCOcolumndensity}.

log($N$(\ce{DCO+})/$N$(DNC)) and log($N$(\ce{DCN})/$N$(DNC)) show strong correlations with kinetic gas temperature. 
Anti-correlation with log($N$(\ce{DCO+})/$N$(DNC)) is found, $r = -0.63$ and a negligible $r = -0.48$ with log($N$(\ce{DCO+})/$N$(DCN)).
On the contrary, a positive correlation is found between log($N$(\ce{DCN})/$N$(DNC)) with a Pearson correlation coefficient of $r = +0.66$. 
It should be noted that the kinetic gas temperature is used to determine the column densities. 
This reflects on the formation pathway dependencies with temperature, where the production of \ce{DCO+} is highest in the lowest temperature, and less for DNC and least for DCN.

\end{appendix}

\end{document}